\documentclass[journal]{IEEEtran} \newcommand{\figsize}{2.85}
\usepackage{graphicx,psfrag}
\usepackage{url}
\usepackage[usenames]{color}
\usepackage[cmex10]{amsmath}
\usepackage{amsfonts,amssymb,latexsym,cite,color,multirow,rotating}
\usepackage{algorithm,algorithmic}

 \newcommand{\putTable}[3]{\begin{table}[t]
  			    \centering
		            #3
     			    \caption{#2}
     			    \label{tab:#1}
			    \vspace{-7mm}
			  \end{table} }
 \newcommand{\putFrag}[4]{\begin{figure}[h!]
                            \centering
                            #4
			    \includegraphics[width=#3in]{figures/#1.eps}
			    \vspace{-2mm}
            		    \caption{#2}
           		    \label{fig:#1}
			    \vspace{-1 mm}
                          \end{figure} }

 \newcommand{\defn}{\triangleq}

 \newcommand{\hvec}[1]{\ensuremath{\Hat{\boldsymbol{#1}}}}
    
 \renewcommand{\vec}[1]{\ensuremath{\boldsymbol{#1}}}

 \newcommand{\norm}[1]{\ensuremath{\| #1 \|}}
 \newcommand{\mc}[1]{\ensuremath{\mathcal{#1}}}

 \newcommand{\Real}{{\mathbb{R}}}
 \newcommand{\Complex}{{\mathbb{C}}}
 \newcommand{\Int}{{\mathbb{Z}}}

 \newcommand{\tran}{^\textsf{T}}
 \newcommand{\herm}{^\textsf{H}}

 \newcommand{\giv}{\,|\,}
 \newcommand{\biggiv}{\,\big|\,}

 \DeclareMathOperator{\real}{Re}

 \DeclareMathOperator{\E}{E}
 \DeclareMathOperator{\var}{var}

 \DeclareMathOperator{\tr}{tr}

 \DeclareMathOperator*{\argmax}{arg\, max}

 \renewcommand{\eqref}[1]{(\ref{eq:#1})}
 
 \newcommand{\Figref}[1]{Figure~\ref{fig:#1}}
 \newcommand{\Figsref}[2]{Figures~\ref{fig:#1}--\ref{fig:#2}}
 \newcommand{\figref}[1]{Fig.~\ref{fig:#1}}
 \newcommand{\figsref}[2]{Figs.~\ref{fig:#1}--\ref{fig:#2}}
 \newcommand{\tabref}[1]{Table~\ref{tab:#1}}
 \newcommand{\secref}[1]{Section~\ref{sec:#1}}
 \newcommand{\Secref}[1]{Section~\ref{sec:#1}}

 \newcounter{comment}[section]
 
 \newcounter{texthead}[section]

 \newcommand{\SNR}{\textsf{SNR}}
 \newcommand{\NMSE}{\textsf{NMSE}}
 
 \newcommand{\TNMSE}{\textsf{TNMSE}}

 \newcommand{\epsball}{\mathcal{B}_\epsilon}
 \newcommand{\epsballC}{\overline{\mathcal{B}_\epsilon}}
 \newcommand{\epseq}{\overset{\epsilon\rightarrow 0}{=}}
 
 \newcommand{\lasso}{_{\textsf{lasso}}}
\newcommand{\Lold}{L^j}
\newcommand{\const}{\text{const}}
\newcommand{\LL}{\mc{L}_{\Lold}}
\newcommand{\bic}{_{\text{\sf BIC}}}

\begin{document}
\setlength{\arraycolsep}{0.4mm}
 \title{Expectation-maximization Gaussian-mixture Approximate Message Passing}
	 \author{Jeremy P. Vila, \emph{Student Member, IEEE}, and Philip Schniter, \emph{Senior Member, IEEE}
	 \thanks{Manuscript received July 12, 2012; revised January 29, 2013; accepted June 19, 2013.  The associate editor coordinating the review of this manuscript and approving it for publication was Prof. Namrata Vaswani.  This work was supported in part by NSF-I/UCRC grant IIP-0968910, by NSF grant CCF-1018368, and by DARPA/ONR grant N66001-10-1-4090.  Portions of this work were presented at the Duke Workshop on Sensing and Analysis of High-Dimensional Data in July 2011 \cite{Vila:Duke:11}; the Asilomar Conference on Signals, Systems, and Computers in Nov.\ 2011 \cite{Vila:ASIL:11}; and the Conference on Information Science and Systems in Mar.\ 2012 \cite{Vila:CISS:12}.}%
	 \thanks{Copyright \copyright 2012 IEEE. Personal use of this material is permitted. However, permission to use this material for any other purposes must be obtained from the IEEE by sending a request to pubs-permissions@ieee.org.}
	 \thanks{The authors are with the Department of Electrical and Computer Engineering, The Ohio State University, Columbus, OH 43210 USA (e-mail: vila.2@osu.edu; schniter@ece.osu.edu).}%
		}
 \date{\today}
 \maketitle

\begin{abstract}
When recovering a sparse signal from noisy compressive linear measurements, the distribution of the signal's non-zero coefficients can have a profound effect on recovery mean-squared error (MSE).
If this distribution was apriori known, then one could use computationally efficient approximate message passing (AMP) techniques for nearly minimum MSE (MMSE) recovery.
In practice, though, the distribution is unknown, motivating the use of robust algorithms like LASSO---which is nearly minimax optimal---at the cost of significantly larger MSE for non-least-favorable distributions.
As an alternative, we propose an empirical-Bayesian technique that simultaneously learns the signal distribution while MMSE-recovering the signal---according to the learned distribution---using AMP.
In particular, we model the non-zero distribution as a Gaussian mixture, and learn its parameters through expectation maximization, using AMP to implement the expectation step. 
Numerical experiments on a wide range of signal classes confirm the state-of-the-art performance of our approach, in both reconstruction error and runtime, in the high-dimensional regime, for most (but not all) sensing operators.
\end{abstract}

%%%%%%%%%%%%%%%%%%%%%%%%%%%%%%%%%%%%%%%%%%%%%%%%%%%%%%%%%%%%%%%%%%%%%%%%%%%%%
\section{Introduction} 				\label{sec:intro}

%\textr{[Note to the publication staff: We have spent a great deal of effort formatting mathematical expressions to maximize readability.  \emph{Please}, if possible, do not increase the size of parentheses, the size of variables, the spacing between math quantities, etc.]}

We consider estimating a $K$-sparse (or compressible) signal $\vec{x} \in \Real^N$ from $M<N$ linear measurements $\vec{y} = \vec{A} \vec{x} + \vec{w} \in \Real^M$, where $\vec{A}$ is known and $\vec{w}$ is additive white Gaussian noise (AWGN).
For this problem, accurate (relative to the noise variance) signal recovery is known to be possible with polynomial-complexity algorithms when $\vec{x}$ is sufficiently sparse and when $\vec{A}$ satisfies certain restricted isometry properties \cite{Eldar:Book:12}, or when $\vec{A}$ is large with i.i.d zero-mean sub-Gaussian entries \cite{Bayati:ISIT:12} as discussed below.

LASSO \cite{Tibshirani:JRSSb:96} (or, equivalently, Basis Pursuit Denoising \cite{Chen:JSC:98}), is a well-known approach to the sparse-signal recovery problem that solves the convex problem
\begin{equation}  \hvec{x}\lasso
  = \arg\min_{\hvec{x}} \norm{\vec{y}-\vec{A}\hvec{x}}_2^2         
	+ \lambda\lasso \norm{\hvec{x}}_1,                      \label{eq:lasso}
\end{equation}
with $\lambda\lasso$ a tuning parameter that trades between the sparsity and measurement-fidelity of the solution. 
When $\vec{A}$ is constructed from i.i.d zero-mean sub-Gaussian entries, the performance of LASSO can be sharply characterized in the large system limit (i.e., as $K,M,N\rightarrow\infty$ with fixed undersampling ratio $M/N$ and sparsity ratio $K/M$)
using the so-called phase transition curve (PTC) \cite{Donoho:Phil:09,Bayati:ISIT:12}.
When the observations are noiseless, the PTC bisects the $M/N$-versus-$K/M$ plane into the region where LASSO reconstructs the signal perfectly (with high probability) and the region where it does not.
(See \figsref{BGPT}{BRPT}.)
When the observations are noisy, the same PTC bisects the plane into the regions where LASSO's noise sensitivity (i.e., the ratio of estimation-error power to measurement-noise power under the worst-case signal distribution) is either finite or infinite \cite{Donoho:NoisyPhase:10}.
An important fact about LASSO's noiseless PTC is that it is invariant to the distribution of the nonzero signal coefficients.
In other words, if the vector $\vec{x}$ is drawn i.i.d from the pdf 
\begin{equation}
p_X(x) = \lambda f_X(x) + (1-\lambda) \delta(x), 	\label{eq:p} 
\end{equation}
where $\delta(\cdot)$ is the Dirac delta, $f_X(\cdot)$ is the active-coefficient pdf (with zero probability mass at $x=0$), and $\lambda\defn K/N$, then the LASSO PTC is invariant to $f_X(\cdot)$.
While this implies that LASSO is robust to ``difficult'' instances of $f_X(\cdot)$, it also implies that LASSO cannot benefit from the case that $f_X(\cdot)$ is an ``easy'' distribution.
For example, when the signal is known apriori to be nonnegative, polynomial-complexity algorithms exist with PTCs that are better than LASSO's \cite{Donoho:PNAS:09}.

At the other end of the spectrum is minimum mean-squared error (MMSE)-optimal signal recovery under \emph{known} marginal pdfs of the form \eqref{p} and \emph{known} noise variance. 
The PTC of MMSE recovery has been recently characterized \cite{Wu:TIT:12} and shown to be well above that of LASSO. 
In particular, for \emph{any} $f_X(\cdot)$, the PTC on the $M/N$-versus-$K/M$ plane reduces to the line $K/M=1$ in both the noiseless and noisy cases.
Moreover, efficient algorithms for approximate MMSE-recovery have been proposed, such as the Bayesian version of
Donoho, Maleki, and Montanari's \emph{approximate message passing} (AMP) algorithm from \cite{Donoho:ITW:10a}, 
which performs loopy belief-propagation on the underlying factor graph using central-limit-theorem approximations that become exact in the large-system limit under i.i.d zero-mean sub-Gaussian $\vec{A}$.
In fact, in this regime, AMP obeys \cite{Bayati:TIT:11} a state-evolution whose fixed points, when unique, are optimal. 
To handle arbitrary noise distributions and a wider class of matrices $\vec{A}$, Rangan proposed a \emph{generalized AMP} (GAMP) \cite{Rangan:ISIT:11} that forms the starting point of this work.
(See \tabref{gamp}.)
For more details and background on GAMP, we refer the reader to \cite{Rangan:ISIT:11}.

In practice, one ideally wants a recovery algorithm that does not need to know $p_X(\cdot)$ and the noise variance a priori, yet offers performance on par with MMSE recovery, which (by definition) requires knowing these prior statistics.
Towards this goal, we propose a recovery scheme that aims to \emph{learn} the prior signal distribution $p_X(\cdot)$, as well as the variance of the AWGN, while simultaneously recovering the signal vector $\vec{x}$ from the noisy compressed measurements $\vec{y}$.
To do so, we model the active component $f_X(\cdot)$ in \eqref{p} using a generic $L$-term Gaussian mixture (GM) and then learn the GM parameters and noise variance using the expectation-maximization (EM) algorithm \cite{Dempster:JRSS:77}.
As we will see, all of the quantities needed for the EM updates are already computed by the GAMP algorithm, making the overall process very computationally efficient.
Moreover, GAMP provides approximately MMSE estimates of $\vec{x}$ that suffice for signal recovery, as well as posterior activity probabilities that suffice for support recovery.

Since, in our approach, the prior pdf parameters are treated as deterministic unknowns, our proposed EM-GM-AMP algorithm can be classified as an ``empirical-Bayesian'' approach \cite{Efron:Book:10}.
Compared with previously proposed empirical-Bayesian approaches to compressive sensing (e.g., \cite{Tipping:JMLR:01,Wipf:TSP:04,Ji:TSP:08}), ours has a more flexible signal model, and thus is able to better match a wide range of signal pdfs $p_X(\cdot)$, as we demonstrate through a detailed numerical study.
In addition, the complexity scaling of our algorithm is superior to that in \cite{Tipping:JMLR:01,Wipf:TSP:04,Ji:TSP:08}, implying lower complexity in the high dimensional regime, as we confirm numerically.
Supplemental experiments demonstrate that our excellent results hold for a wide range of sensing operators $\vec{A}$, with some exceptions.
Although this paper does not contain any convergence guarantees or a rigorous analysis/justification of the proposed EM-GM-AMP, Kamilov et al.\ showed (after the submission of this work) in \cite{Kamilov:NIPS:12} that a generalization of EM-GM-AMP yields asymptotically (i.e., in the large system limit) consistent parameter estimates when $\vec{A}$ is i.i.d zero-mean Gaussian, when the parameterized signal and noise distributions match the true signal and noise distributions, and when those distributions satisfy certain identifiability conditions. 
We refer interested readers to \cite{Kamilov:NIPS:12} for more details.

\emph{Notation}: 
  For matrices, we use boldface capital letters like $\vec{A}$, 
  and we use $\tr(\vec{A})$ and $\norm{\vec{A}}_F$ to denote 
  the trace and Frobenius norm, respectively.
  Moreover, we use
  $(\cdot)\tran$, $(\cdot)^*$, and $(\cdot)\herm$ to denote
  transpose, conjugate, and conjugate transpose, respectively.
  For vectors, we use boldface small letters like $\vec{x}$, and we use 
  $\norm{\vec{x}}_p=(\sum_n |x_n|^p)^{1/p}$ to denote the $\ell_p$ norm, 
  with $x_n$ representing the $n^{th}$ element of $\vec{x}$.
  For a Gaussian random vector $\vec{x}$ with mean $\vec{m}$ and
  covariance matrix $\vec{Q}$, we denote the pdf by
  $\mc{N}(\vec{x};\vec{m},\vec{Q})$, and for its 
  circular complex Gaussian counterpart, we use 
  $\mc{CN}(\vec{x};\vec{m},\vec{Q})$.
  Finally,
  we use $\E\{\cdot\}$, $\delta(\cdot)$, $\Real$, and $\Complex$  
  to denote the expectation operation,
  the Dirac delta,
  the real field, 
  and the complex field, respectively.

%%%%%%%%%%%%%%%%%%%%%%%%%%%%%%%%%%%%%%%%%%%%%%%%%%%%%%%%%%%%%%%%%%%%%%%%%%%%%
\section{Gaussian-Mixture GAMP}	\label{sec:GMAMP}

We first introduce Gaussian-mixture (GM) GAMP, a key component of our overall approach, 
where the coefficients in $\vec{x}=[x_1,\dots,x_N]\tran$ are assumed to be i.i.d with marginal pdf
\begin{equation}
  p_X(x;\lambda,\vec{\omega},\vec{\theta},\vec{\phi})
  = (1-\lambda) \delta(x) +  \lambda \sum_{\ell = 1}^L \omega_\ell \mc{N}  (x;\theta_\ell,\phi_\ell) ,
  					\label{eq:pX}
\end{equation}
where $\delta(\cdot)$ is the Dirac delta,
$\lambda$ is the sparsity rate, and,
for the $k^{th}$ GM component, $\omega_k$, $\theta_k$, and $\phi_k$
are the weight, mean, and variance, respectively.
In the sequel, we use $\vec{\omega}\defn[\omega_1,\dots,\omega_L]\tran$ 
and similar definitions for $\vec{\theta}$ and $\vec{\phi}$.
By definition, $\sum_{\ell=1}^L \omega_\ell = 1$.
The noise $\vec{w}=[w_1,\dots,w_M]\tran$ is assumed to be i.i.d
Gaussian,
with mean zero and variance $\psi$, i.e.,
\begin{equation}
  p_W(w;\psi)
  = \mc{N}(w;0,\psi) ,		\label{eq:pW}
\end{equation}
and independent of $\vec{x}$.
Although above and in the sequel we assume real-valued quantities, all expressions in the sequel can be converted to the circular-complex case by replacing $\mc{N}$ with $\mc{CN}$ 
and removing the $\frac{1}{2}$'s from \eqref{psi2}, \eqref{phideriv}, and \eqref{phikder}.
We note that, from the perspective of GM-GAMP, the prior parameters $\vec{q}\defn[\lambda,\vec{\omega},\vec{\theta},\vec{\phi},\psi]$ and the number of mixture components, $L$, are treated as fixed and known.

GAMP models the relationship between the $m^{th}$ observed output $y_m$ and the corresponding noiseless output $z_m\defn \vec{a}_m\tran \vec{x}$, where $\vec{a}_m\tran$ denotes the $m^{th}$ row of $\vec{A}$, using the conditional pdf $p_{Y|Z}(y_m|z_m;\vec{q})$.
It then approximates the true marginal posterior $p(z_m|\vec{y};\vec{q})$ by
\begin{equation}
  p_{Z|\vec{Y}}(z_m|\vec{y};\hat{p}_m, \mu^p_m,\vec{q})
  \defn 
  \frac{ p_{Y|Z}(y_m|z_m;\vec{q}) \,\mc{N}(z_m;\hat{p}_m,\mu^p_m) }
  { \int_z p_{Y|Z}(y_m|z;\vec{q}) \,\mc{N}(z;\hat{p}_m,\mu^p_m) }
  	\label{eq:gamppostz}
\end{equation}
using quantities $\hat{p}_m$ and $\mu^p_m$ that change with iteration $t$ (see \tabref{gamp}), although here we suppress the $t$ notation for brevity.
Under the AWGN assumption\footnote{%
Because GAMP can handle an arbitrary $p_{Y|Z}(\cdot|\cdot)$, the extension of EM-GM-AMP to additive non-Gaussian noise, and even non-additive measurement channels (such as with quantized outputs \cite{Kamilov:SPL:12} or logistic regression \cite{Rangan:ISIT:11}), is straightforward. 
Moreover, the parameters of the pdf $p_{Y|Z}(\cdot|\cdot)$ could be learned using a method similar to that which we propose for learning the AWGN variance $\psi$, as will be evident from the derivation in \secref{psiest}.  Finally, one could even model $p_{Y|Z}(\cdot|\cdot)$ as a Gaussian mixture and learn the corresponding parameters.}
\eqref{pW} we have $p_{Y|Z}(y|z;\vec{q})=\mc{N}(y;z,\psi)$, 
and thus the pdf \eqref{gamppostz} has moments \cite{Rangan:ISIT:11}
\begin{align}
 E_{Z|\vec{Y}}\{z_m|\vec{y};\hat{p}_m, \mu^p_m,\vec{q}\} 
 &= \hat{p}_m + \tfrac{\mu^p_m}{\mu^p_m+\psi}(y_m - \hat{p}_m) 
 	\label{eq:zhat}\\
 \var_{Z|\vec{Y}}\{z_m|\vec{y};\hat{p}_m, \mu^p_m,\vec{q}\} 
 &= \frac{\mu^p_m\psi}{\mu^p_m + \psi}.
 	\label{eq:zvar}
\end{align}
GAMP then approximates the true marginal posterior $p(x_n|\vec{y};\vec{q})$ by
\begin{equation}
 p_{X|\vec{Y}}(x_n|\vec{y} ; \hat{r}_n, \mu^r_n,\vec{q}) 
 \defn \frac{ p_X(x_n; \vec{q}) \, \mc{N}(x_n; \hat{r}_n, \mu^r_n) }
 	{ \int_{x}p_X(x; \vec{q}) \, \mc{N}(x; \hat{r}_n, \mu^r_n) }
	\label{eq:gamppost}
\end{equation}
where again $\hat{r}_n$ and $\mu^r_n$ vary with the GAMP iteration $t$.

Plugging the sparse GM prior \eqref{pX} into \eqref{gamppost} and simplifying, one can obtain% 
\footnote{Both \eqref{GAMPpost2} and \eqref{zeta} can be derived from \eqref{GAMPpost} via the Gaussian-pdf multiplication rule:
  $\mc{N}(x;a,\!A)\mc{N}(x;b,\!B) 
  \!=\! \mc{N}(x;\frac{a/A + b/B}{1/A+1/B},\frac{1}{1/A+1/B})
  \mc{N}(0; a-b, A+B)$.}
the GM-GAMP approximated posterior
\begin{eqnarray}
 \lefteqn{ p_{X|\vec{Y}}(x_n|\vec{y};\hat{r}_n,\mu^r_n,\vec{q}) 
 }\nonumber\\
 &=& 
 	\bigg( \!(1\!-\!\lambda)\delta(x_n) 
 	\!+\! \lambda\sum_{\ell=1}^L \omega_\ell \mc{N}(x_n;\theta_\ell,\phi_\ell) \!\bigg) \frac{\mc{N}(x_n; \hat{r}_n, \mu^r_n) }{ \zeta_n }  \quad
			\label{eq:GAMPpost} \\
 &=& \big(1-\pi_n\big)\delta(x_n) + \pi_n \sum_{\ell=1}^L \overline{\beta}_{n,\ell}\, 
 	\mc{N}\big(x_n;\gamma_{n,\ell},\nu_{n,\ell}\big)
			\label{eq:GAMPpost2} 
\end{eqnarray}
with normalization factor 
\begin{eqnarray}
 &&\zeta_n
 \defn \int_x p_X(x; \vec{q}) \, \mc{N}(x; \hat{r}_n, \mu^r_n)  \\
 &&= (1\!-\!\lambda)\mc{N}(0;\hat{r}_n,\mu^r_n) 
 	\!+\! \lambda \sum_{\ell=1}^L \omega_\ell \mc{N}(0;\hat{r}_n\!-\!\theta_\ell,\mu^r_n\!+\!\phi_\ell) \qquad \label{eq:zeta} 
\end{eqnarray}
and $(\hat{r}_n,\mu^r_n,\vec{q})$-dependent quantities
\begin{align}
\beta_{n,\ell}
&\defn 
\lambda \omega_\ell \mc{N}(\hat{r}_n;\theta_\ell,\phi_\ell + \mu^r_n)
	\label{eq:beta}\\
\overline{\beta}_{n,\ell}
&\defn 
\frac{\beta_{n,\ell}}{\sum_{k=1}^L \beta_{n,k}}
	\label{eq:betabar}\\
\pi_n  
  &\defn \frac{1}{1+\left(\frac{\sum_{\ell=1}^L \beta_{n,\ell}}{(1-\lambda)\mc{N}(0;\hat{r}_n,\mu^r_n)}\right)^{-1}}
  	\label{eq:pi}\\
\gamma_{n,\ell} 
&\defn 
\frac{\hat{r}_n/\mu^r_n+\theta_\ell/\phi_\ell}{1/\mu^r_n + 1/\phi_\ell}
	\label{eq:gamma}\\
\nu_{n,\ell}
&\defn \frac{1}{1/\mu^r_n + 1/\phi_\ell} .			\label{eq:nu}
\end{align}
The posterior mean and variance of $p_{X|\vec{Y}}$ are given in steps (R9)-(R10) of \tabref{gamp}, and
\eqref{GAMPpost2} makes it clear that $\pi_n$ is GM-GAMP's approximation of the posterior support probability $\Pr\{x_n\!\neq\! 0\giv\vec{y};\vec{q}\}$.

In principle, one could specify GAMP for an arbitrary signal prior $p_X(\cdot)$.
However, if the integrals in (R9)--(R10) are not computable in closed form (e.g., when $p_X(\cdot)$ is Student's-t), then they would need to be computed numerically, thereby drastically increasing the computational complexity of GAMP. 
In contrast, for GM signal models, we see above that all steps can be computed in closed form.
Thus, a practical approach to the use of GAMP with an intractable signal prior $p_X(\cdot)$ is to \emph{approximate} $p_X(\cdot)$ using an $L$-term GM, after which all GAMP steps can be easily implemented.
The same approach could also be used to ease the implementation of intractable output priors $p_{Y|Z}(\cdot|\cdot)$.

\putTable{gamp}{The GAMP Algorithm from \cite{Rangan:ISIT:11} with a stopping condition in (R10) that uses the normalized tolerance parameter $\tau_{\textsf{gamp}}$}{\footnotesize
\begin{equation*}
\begin{array}{|lrcl@{}r|}\hline
  \multicolumn{4}{|l}{\textsf{inputs:~~}
  	p_{X}(\cdot), 
	p_{Y|Z}(\cdot|\cdot), 
	\{A_{mn}\}, T_{\max}, \tau_{\textsf{gamp}} 
	}&\\[1mm]
  \multicolumn{2}{|l}{\textsf{definitions:}}&&&\\[-1mm]
  &p_{Z|\vec{Y}}(z_m|\vec{y};\hat{p}_m,\mu^p_m,\vec{q})
   &\defn& \frac{p_{Y|Z}(y_m|z_m;\vec{q}) \,\mc{N}(z_m;\hat{p}_m,\mu^p_m)}
	{\int_{z} p_{Y|Z}(y_m|z;\vec{q}) \,\mc{N}(z;\hat{p}_m,\mu^p_m)} &\text{(D1)}\\
  &p_{X|\vec{Y}}\!(x_n|\vec{y};\hat{r}_n,\mu^r_n,\vec{q})
   &\defn& \frac{p_{X}\!(x_n;\vec{q}) \,\mc{N}(x_n;\hat{r}_n,\mu^r_n)}
        {\int_{x}p_{X}\!(x;\vec{q}) \,\mc{N}(x;\hat{r}_n,\mu^r_n)}&\text{(D2)}\\
  \multicolumn{2}{|l}{\textsf{initialize:}}&&&\\
  &\forall n: 
   \hat{x}_{n}(1) &=& \int_{x} x\, p_{X}(x) & \text{(I1)}\\
  &\forall n:
   \mu^x_n(1) &=& \int_{x} |x-\hat{x}_n(1)|^2  p_{X}(x) & \text{(I2)}\\
  &\forall m: 
   \hat{s}_{m}(0) &=& 0 & \text{(I3)}\\
  \multicolumn{2}{|l}{\textsf{for $t=1:T_{\max}$,}}&&&\\
  &\forall m:
   \mu^p_m(t)
   &=& \textstyle \sum_{n=1}^{N} |\!A_{mn}|^2 \mu^x_{n}(t) & \text{(R1)}\\
  &\forall m:
   \hat{p}_m(t)
   &=& \sum_{n=1}^{N} \!A_{mn} \hat{x}_{n}(t) - \mu^p_m(t) \,\hat{s}_m(t-1)& \text{(R2)}\\
  &\forall m:
   \mu^z_m(t)
   &=& \var_{Z|\vec{Y}}\{z_m|\vec{y};\hat{p}_m(t),\mu^p_m(t),\vec{q}\} & \text{(R3)}\\
  &\forall m:
   \hat{z}_m(t)
   &=& \E_{Z|\vec{Y}}\{z_m|\vec{y};\hat{p}_m(t),\mu^p_m(t),\vec{q}\} & \text{(R4)}\\
  &\forall m:
   \mu^s_m(t)
   &=& \big(1-\mu^z_m(t)/\mu^p_m(t)\big)/\mu^p_m(t) & \text{(R5)}\\
  &\forall m:
   \hat{s}_m(t)
   &=& \big(\hat{z}_m(t)-\hat{p}_m(t)\big)/\mu^p_m(t) & \text{(R6)}\\
  &\forall n:
   \mu^r_n(t)
   &=& \textstyle \big(\sum_{m=1}^{M} |\!A_{mn}|^2 \mu^s_m(t) 
	\big)^{-1} & \text{(R7)}\\
  &\forall n:
   \hat{r}_n(t)
   &=& \textstyle \hat{x}_n(t)+ \mu^r_n(t) \sum_{m=1}^{M} \!A_{mn}^*
	\hat{s}_{m}(t)  & \text{(R8)}\\
  &\forall n:
   \mu^x_n(t\!+\!1)
   &=& \var_{X|\vec{Y}}\{x_n|\vec{y};\hat{r}_n(t),\mu^r_n(t),\vec{q}\} & \text{(R9)}\\
  &\forall n:
   \hat{x}_{n}(t\!+\!1)
   &=& \E_{X|\vec{Y}}\{x_n|\vec{y};\hat{r}_n(t),\mu^r_n(t),\vec{q}\} & \text{(R10)}\\
  &\multicolumn{3}{l}{\hspace{3.5mm}
      \textsf{if} \sum_{n=1}^N|\hat{x}_n(t\!+\!1)-\hat{x}_n(t)|^2 
  	< \tau_{\textsf{gamp}} \sum_{n=1}^N|\hat{x}_n(t)|^2,
      \textsf{break}} & \text{(R11)}\\
  \multicolumn{2}{|l}{\textsf{end}}&&&\\[1mm]
  \multicolumn{4}{|l}{\textsf{outputs:~~}
	\{\hat{z}_m(t),\mu^z_m(t)\},
	\{\hat{r}_n(t),\mu^r_n(t)\}, 
	\{\hat{x}_n(t\!+\!1),\mu^x_n(t\!+\!1)\} 
	}&\\[1mm]
  \hline
\end{array}
\end{equation*}
\vspace{-4mm}
}

%%%%%%%%%%%%%%%%%%%%%%%%%%%%%%%%%%%%%%%%%%%%%%%%%%%%%%%%%%%%%%%%%%%%%%%%
\section{EM Learning of the Prior Parameters $\vec{q}$}	\label{sec:EMalg}

We now propose an expectation-maximization (EM) algorithm \cite{Dempster:JRSS:77} to learn the prior parameters $\vec{q}\defn[\lambda,\vec{\omega},\vec{\theta},\vec{\phi},\psi]$.
The EM algorithm is an iterative technique that increases a lower bound on the likelihood $p(\vec{y};\vec{q})$ at each iteration, thus guaranteeing that the likelihood converges to a local maximum or at least a saddle point \cite{Wu:AS:83}.
In our case, 
the EM algorithm manifests as follows.
Writing, for arbitrary pdf $\hat{p}(\vec{x})$,
\begin{align}
&\ln p(\vec{y};\vec{q}) 
=\int_{\vec{x}} \hat{p}(\vec{x}) \ln p(\vec{y};\vec{q}) \\
&=\int_{\vec{x}} \hat{p}(\vec{x}) \ln \Big( 
	\frac{p(\vec{x},\vec{y};\vec{q})}{\hat{p}(\vec{x})}
	\frac{\hat{p}(\vec{x})}{p(\vec{x}|\vec{y};\vec{q})} \Big)\\
&=\underbrace{ E_{\hat{p}(\vec{x})}\{\ln p(\vec{x},\vec{y};\vec{q})\}
  + H(\hat{p}) }_{\displaystyle \defn \mc{L}_{\hat{p}}(\vec{y};\vec{q})}
  + \underbrace{ D(\hat{p}\,\|\,p_{\vec{X}|\vec{Y}}(\cdot|\vec{y};\vec{q})) 
  	}_{\displaystyle \geq 0} 
\end{align}
where 
$\E_{\hat{p}(\vec{x})}\{\cdot\}$ denotes expectation over $\vec{x}\!\sim\! \hat{p}(\vec{x})$, 
$H(\hat{p})$ denotes the entropy of pdf $\hat{p}$, and 
$D(\hat{p}\,\|\,p)$ denotes the Kullback-Leibler (KL) divergence between $\hat{p}$ and $p$. The non-negativity of the KL divergence implies that $\mc{L}_{\hat{p}}(\vec{y};\vec{q})$ is a lower bound on $\ln p(\vec{y};\vec{q})$, and thus
the EM algorithm iterates over two steps: E) choosing $\hat{p}$ to maximize the lower bound for fixed $\vec{q}\!=\!\vec{q}^i$, and M) choosing $\vec{q}$ to maximize the lower bound for fixed $\hat{p}\!=\!\hat{p}^i$.
For the E step, since $\mc{L}_{\hat{p}}(\vec{y};\vec{q}^i) = \ln p(\vec{y};\vec{q}^i) - D(\hat{p}\,\|\,p_{\vec{X}|\vec{Y}}(\cdot|\vec{y};\vec{q}^i))$, the maximizing pdf would clearly be $\hat{p}^i(\vec{x})=p_{\vec{X}|\vec{Y}}(\vec{x}|\vec{y};\vec{q}^i)$, i.e., the true posterior under prior parameters $\vec{q}^i$. 
Then, for the M step, since $\mc{L}_{\hat{p}^i}(\vec{y};\vec{q})=\E_{\hat{p}^i(\vec{x})}\{\ln p(\vec{x},\vec{y};\vec{q})\}+H(\hat{p}^i)$, the maximizing $\vec{q}$ would clearly be
$\vec{q}^{i+1}=\argmax_{\vec{q}} \E\{\ln p(\vec{x},\vec{y};\vec{q})\giv\vec{y};\vec{q}^i\}$.

In our case, because the true posterior is very difficult to calculate, we instead construct our lower-bound $\mc{L}_{\hat{p}}(\vec{y};\vec{q})$ using the GAMP approximated posteriors, i.e., we set $\hat{p}^i(\vec{x})\!=\!\prod_n p_{X|\vec{Y}}(x_n|\vec{y};\vec{q}^i)$ for $p_{X|\vec{Y}}$ defined in \eqref{gamppost},
resulting in
\begin{align}
\vec{q}^{i+1}
&=\argmax_{\vec{q}} \hat{\E}\{\ln p(\vec{x},\vec{y};\vec{q})\giv \vec{y};\vec{q}^i\},
\label{eq:EMmain}
\end{align}
where ``$\hat{\E}$'' indicates the use of the GAMP's posterior approximation.
Moreover, since the joint optimization in \eqref{EMmain} is difficult to perform, 
we update $\vec{q}$ one component at a time (while holding the others fixed), which is the well known ``incremental'' variant on EM from \cite{Neal:Jordan:99}.
In the sequel, we use ``$\vec{q}^i_{\setminus \lambda}$'' to denote the vector $\vec{q}^i$ 
with the element $\lambda$ removed (and similar for the other parameters).

%%%%%%%%%%%%%%%%%%%%%%%%%%%%%%%%%%%%%%%%%%%%%%%%%%%%%%%%%%%%%%%%%%%%%
\subsection{EM Update of the Gaussian Noise Variance $\psi$}		\label{sec:psiest}

We first derive the EM update for the noise variance $\psi$ given a previous parameter estimate $\vec{q}^i$.
For this, we write 
$p(\vec{x},\vec{y};\vec{q})=C p(\vec{y}|\vec{x};\psi)=C\prod_{m=1}^M p_{Y|Z}(y_m|\vec{a}_m\tran\vec{x};\psi)$ 
for a $\psi$-invariant constant $C$, so that
\begin{eqnarray}
\lefteqn{ \psi^{i+1} 
= \argmax_{\psi>0} \sum_{m=1}^M \hat{\E}\big\{\ln p_{Y|Z}(y_m|\vec{a}_m\tran\vec{x};\psi) \biggiv \vec{y};\vec{q}^i\big\} }\\
&=& \argmax_{\psi>0} \sum_{m=1}^M \int_{z_m} \!\!p_{Z|\vec{Y}}(z_m|\vec{y};\vec{q}^i) \ln p_{Y|Z}(y_m|z_m;\psi)  \qquad
\label{eq:EMpsi}
\end{eqnarray}
since $z_m=\vec{a}_m\tran\vec{x}$.
The maximizing value of $\psi$ in \eqref{EMpsi} is necessarily a value of
$\psi$ that zeroes the derivative of the sum, i.e., that satisfies\footnote{The continuity of both the integrand and its partial derivative with respect to $\psi$ allow the use of Leibniz's integral rule to exchange differentiation and integration.}
\begin{equation}
 \sum_{m=1}^M \int_{z_m} p_{Z|\vec{Y}}(z_m|\vec{y};\vec{q}^i)
 	\frac{d}{d\psi}\ln p_{Y|Z}(y_m|z_m;\psi) = 0.
 \label{eq:GAMPpsi}
\end{equation}
Because $p_{Y|Z}(y_m|z_m;\psi)=\mc{N}(y_m;z_m,\psi)$, we can obtain
\begin{equation}
 \frac{d}{d\psi}\ln p_{Y|Z}(y_m|z_m;\psi) 
 = \frac{1}{2}\left( \frac{|y_m-z_m|^2}{\psi^2} - \frac{1}{\psi} \right),
 \label{eq:psi2}
\end{equation}
which, when plugged into \eqref{GAMPpsi}, yields the unique solution
\begin{align}
 \psi^{i+1}
 &= \frac{1}{M}\sum_{m=1}^M \int_{z_m} p_{Z|\vec{Y}}(z_m|\vec{y};\vec{q}^i)\, |y_m-z_m|^2 \\
 &= \frac{1}{M}\sum_{m=1}^M \big(|y_m-\hat{z}_m|^2 + \mu^z_m\}\big) ,
 \label{eq:EMpsi2}
\end{align}
where the use of $\hat{z}_m$ and $\mu^z_m$ follows from (R3)-(R4) in \tabref{gamp}.
\vspace{-5mm}

%%%%%%%%%%%%%%%%%%%%%%%%%%%%%%%%%%%%%%%%%%%%%%%%%%%%%%%%%%%%%%%%%%%%%%%%
\subsection{EM Updates of the Signal Parameters: BG Case} \label{sec:EMBG}
Suppose that the signal distribution $p_X(\cdot)$ is modeled using an $L\!=\!1$-term GM, i.e., a Bernoulli-Gaussian (BG) pdf. 
In this case, the marginal signal prior in \eqref{pX} reduces to
\begin{equation}
  p_X(x;\lambda,\omega,\theta,\phi)
  = (1-\lambda) \delta(x) +  \lambda\mc{N}  (x;\theta,\phi).
  					\label{eq:pXBG}
\end{equation}
Note that, in the BG case, the mixture weight $\omega$ is, by definition, unity and does not need to be learned.  

We now derive the EM update for $\lambda$ given previous parameters
$\vec{q}^i \defn [\lambda^i,\theta^i,\phi^i,\psi^i]$.
Because we can write $p(\vec{x},\vec{y};\vec{q})
=C\prod_{n=1}^N p_X(x_n;\lambda,\theta,\phi)$ for 
a $\lambda$-invariant constant $C$,
\begin{equation}
  \lambda^{i+1} 
  = \argmax_{\lambda\in(0,1)} \sum_{n=1}^N 
  	\hat{\E}\big\{\ln p_X(x_n;\lambda,\vec{q}^i_{\setminus\lambda})\biggiv \vec{y};\vec{q}^i\big\}.
\label{eq:EMlamBG}
\end{equation}
The maximizing value of $\lambda$ in \eqref{EMlamBG} is necessarily a value of $\lambda$ that zeroes the derivative of the sum, i.e., that satisfies\footnote{To justify the exchange of differentiation and integration via Leibniz's integral rule here, one could employ the Dirac approximation $\delta(x)=\mc{N}(x;0,\varepsilon)$ for fixed arbitrarily small $\varepsilon>0$, after which the integrand and its derivative w.r.t $\lambda$ become continuous.  The same comment applies in to all exchanges of differentiation and integration in the sequel.}
\begin{eqnarray}
 \sum_{n=1}^N \int_{x_n} p_{X|\vec{Y}}(x_n|\vec{y};\vec{q}^i)\frac{d}{d\lambda}\ln p_X(x_n;\lambda,\vec{q}^i_{\setminus\lambda}) = 0. 
 \label{eq:GAMPlamBG}
\end{eqnarray}
For the BG $p_X(x_n;\lambda,\theta,\phi)$ in \eqref{pXBG},
it is readily seen that
\begin{align}
\frac{d}{d\lambda} \ln p_X(x_n;\lambda,\vec{q}^i_{\setminus\lambda}) 
&= \frac{\mc{N}(x_n;\theta^i,\phi^i)-\delta(x_n)}{p_X(x_n;\lambda,\vec{q}^i_{\setminus\lambda})}   \\
&=
\begin{cases}
\frac{1}{\lambda} &x_n \neq 0 \\
\frac{-1}{1-\lambda} &x_n = 0.
\end{cases}
\label{eq:lamderivBG}
\end{align}

Plugging \eqref{lamderivBG} and \eqref{GAMPpost} into \eqref{GAMPlamBG}, it 
becomes evident that the neighborhood around the point $x_n=0$ should be 
treated differently than the remainder of $\Real$.
Thus, we define the closed ball $\epsball\defn[-\epsilon,\epsilon]$ and
its complement $\epsballC \defn \Real\setminus\mc{B}_\epsilon$, and note that,
in the limit $\epsilon\rightarrow 0$, the following is
equivalent to \eqref{GAMPlamBG}:
\begin{equation}
  \sum_{n=1}^N 
  \underbrace{\int_{x_n\in\epsballC} \!\!p_{X|\vec{Y}}(x_n|\vec{y};\vec{q}^i)}_{
  	\displaystyle \epseq\pi_n}
  =\frac{\lambda}{1\!-\!\lambda}\sum_{n=1}^N 
  \underbrace{\int_{x_n\in\epsball} \!\!\!\!p_{X|\vec{Y}}(x_n|\vec{y};\vec{q}^i)}_{
  	\displaystyle \epseq 1\!-\!\pi_n} 
		\label{eq:GAMPlam2BG}
\end{equation}
where the values taken by the integrals are evident from \eqref{GAMPpost2}.
Finally, the EM update for $\lambda$ is the unique value satisfying 
\eqref{GAMPlam2BG} as $\epsilon\rightarrow 0$, which is readily shown to be
\begin{equation}
\lambda^{i+1} 
= \frac{1}{N}\sum_{n=1}^N \pi_n.  
\label{eq:lamresult}
\end{equation}
Conveniently, the posterior support probabilities $\{\pi_n\}_{n=1}^N$ are easily calculated from the GM-GAMP outputs via \eqref{pi}.

Similar to \eqref{EMlamBG}, the EM update for $\theta$ can be written as
\begin{equation}
\theta^{i+1} 
= \argmax_{\theta\in\Real} \sum_{n=1}^N \hat{\E}\big\{\ln p_X(x_n;\theta, \vec{q}^i_{\setminus\theta})\biggiv\vec{y};\vec{q}^i\big\} . 
\label{eq:EMtheta}
\end{equation}
The maximizing value of $\theta$ in \eqref{EMtheta} is again a necessarily a value of
$\theta$ that zeroes the derivative, i.e., that satisfies
\begin{equation}
  \sum_{n=1}^N \int_{x_n} p_{X|\vec{Y}}(x_n|\vec{y};\vec{q}^i)\frac{d}{d\theta}\ln p_X(x_n;\theta,  \vec{q}^i_{\setminus\theta}) = 0 . 
  \label{eq:GAMPtheta} 
\end{equation}
For the BG $p_X(x_n;\lambda,\theta,\phi)$ given in \eqref{pXBG},
\begin{eqnarray}
\frac{d}{d\theta}\ln p_X(x_n;\lambda^i,\theta,\phi^i)
&=& \frac{\left(x_n-\theta\right)}{\phi^i}\frac{\lambda^i\mc{N}(x_n;\theta,\phi^i)}{p_X(x_n;\theta,  \vec{q}^i_{\setminus\theta})} \\
&=& \begin{cases}
\frac{x_n-\theta}{\phi^i} &x_n \neq 0\\
0 &x_n = 0.
\end{cases}
  \label{eq:thetaderiv}
\end{eqnarray}
Splitting the domain of integration in \eqref{GAMPtheta} into $\epsball$ and $\epsballC$ as before, and then plugging in \eqref{thetaderiv}, we find that the following
is equivalent to \eqref{GAMPtheta} in the limit of $\epsilon\rightarrow 0$:
\begin{equation}
 \sum_{n=1}^N \int_{x_n\in\epsballC} (x_n-\theta) \,
 	p_{X|\vec{Y}}(x_n|\vec{y}; \vec{q}^i) = 0.
\label{eq:GAMPtheta2}
\end{equation}
The unique value of $\theta$ satisfying \eqref{GAMPtheta2} as $\epsilon\rightarrow 0$ is then
\begin{align}
 \theta^{i+1} 
 &= \frac{\sum_{n=1}^N \lim_{\epsilon\rightarrow 0}
    	\int_{x_n\in\epsballC} x_n \,p_{X|\vec{Y}}(x_n|\vec{y}; \vec{q}^i)}
 	{\sum_{n=1}^N \lim_{\epsilon\rightarrow 0}
	\int_{x_n\in\epsballC} p_{X|\vec{Y}}(x_n|\vec{y}; \vec{q}^i)} 
				\label{eq:EMtheta2}\\
 &= \frac{1}{\lambda^{i+1} N} \sum_{n=1}^N \pi_n \gamma_{n,1}
				\label{eq:EMtheta3}
\end{align}
where $\{\gamma_{n,1}\}_{n=1}^N$ defined in \eqref{gamma} are easily computed from the GM-GAMP outputs.
The equality in \eqref{EMtheta3} can be verified by plugging the GAMP posterior expression \eqref{GAMPpost2} into \eqref{EMtheta2}.

Similar to \eqref{EMlamBG}, the EM update for $\phi$ can be written as
\begin{equation}
 \hat{\phi}^{i+1} 
 = \argmax_{\phi>0} \sum_{n=1}^N 
 	\hat{\E}\big\{\ln p_X(x_n;\phi,  \vec{q}^i_{\setminus\phi})
	\biggiv \vec{y};\vec{q}^i\big\}.
\label{eq:EMphi}
\end{equation}
The maximizing value of $\phi$ in \eqref{EMphi} is again necessarily a value of $\phi$ that zeroes the derivative, i.e., that satisfies
\begin{equation}
  \sum_{n=1}^N \int_{x_n} p_{X|\vec{Y}}(x_n|\vec{y};\vec{q}^i)
  	\frac{d}{d\phi}\ln p_X(x_n;\phi, \vec{q}^i_{\setminus\phi}) 
	= 0. 
	\label{eq:GAMPphi} 
\end{equation}
For the $p_X(x_n;\lambda,\theta,\phi)$ given in \eqref{pXBG}, it is readily seen that
\begin{eqnarray}
\lefteqn{ \frac{d}{d\phi}\ln p_X(x_n;\lambda^i,\theta^i,\phi) }\nonumber\\
&=& \frac{1}{2}\left(\frac{|x_n-\theta^i|^2}{(\phi)^2}-\frac{1}{\phi}\right)
  \frac{\lambda^i\,\mc{N}(x_n;\theta^i ,\phi)}
       {p_X(x_n;,\phi, \vec{q}^i_{\setminus\phi})} \nonumber \\
 &=& \begin{cases}
 	\frac{1}{2}\left(\frac{|x_n-\theta^i|^2}{(\phi)^2}-\frac{1}{\phi}\right) & x_n \neq 0 \\
       	0 & x_n = 0
     \end{cases}.
	\label{eq:phideriv}
\end{eqnarray}
Splitting the domain of integration in \eqref{GAMPphi} into $\epsball$ and 
$\epsballC$ as before, and then plugging in \eqref{phideriv}, we find that the following
is equivalent to \eqref{GAMPphi} in the limit of $\epsilon\rightarrow 0$:
\begin{equation}
 \sum_{n=1}^N \int_{x_n\in\epsballC} \big(|x_n-\theta^i|^2-\phi\big) 
 	\, p_{X|\vec{Y}}(x_n|\vec{y};\vec{q}^i)= 0.
 \label{eq:GAMPphi2}
\end{equation}
The unique value of $\phi$ satisfying \eqref{GAMPphi2} as $\epsilon\rightarrow 0$ is then
\begin{eqnarray}
 \phi^{i+1} 
 &=& \frac{\sum_{n=1}^N \lim_{\epsilon\rightarrow 0}
    	\int_{x_n\in\epsballC} |x_n-\theta^i|^2 p_{X|\vec{Y}}(x_n|\vec{y}; \vec{q}^i)}
 	{\sum_{n=1}^N \lim_{\epsilon\rightarrow 0}
	\int_{x_n\in\epsballC} p_{X|\vec{Y}}(x_n|\vec{y}; \vec{q}^i)}  .\qquad
				\label{eq:EMphi2}
\end{eqnarray}
Finally, we expand $|x_n - \theta^i|^2 = |x_n|^2 - 2 \real(x_n^* \theta^i) + |\theta^i|^2$ which gives
\begin{equation}
\phi^{i+1} 
 = \frac{1}{\lambda^{i+1} N} \sum_{n=1}^N \pi_n
	\Big( \big|\theta^i-\gamma_{n,1}\big|^2 + \nu_{n,1}
	\Big)
				\label{eq:EMphi3}
\end{equation}
where $\{\nu_{n,1}\}_{n=1}^N$ from \eqref{nu} are easily computed from the GAMP outputs.
The equality in \eqref{EMphi3} can be readily verified by plugging \eqref{GAMPpost2} into \eqref{EMphi2}.

%%%%%%%%%%%%%%%%%%%%%%%%%%%%%%%%%%%%%%%%%%%%%%%%%%%%%%%%%%%%%%%%%%%%%%%%
\subsection{EM Updates of the Signal Parameters: GM Case} \label{sec:EMGM}
We now generalize the EM updates derived in \secref{EMBG} to the GM prior given in \eqref{pX} for $L \geq 1$.
As we shall see, it is not possible to write the exact EM updates in closed-form when $L>1$, and so some approximations will be made.

We begin by deriving the EM update for $\lambda$ given the previous parameters
$\vec{q}^i\defn [\lambda^i,\vec{\omega}^i,\vec{\theta}^i,\vec{\phi}^i,\psi^i]$.
The first two steps are identical to the steps \eqref{EMlamBG} and \eqref{GAMPlamBG} presented for the BG case, and for brevity we do not repeat them here.
In the third step, use of the GM prior \eqref{pX} yields
\begin{align}
 \frac{d}{d\lambda} \ln p_X(x_n;\lambda,\vec{q}^i_{\setminus\lambda}) 
&= \frac{\sum_{\ell=1}^L \omega^i_\ell \mc{N}(x_n;\theta_\ell^i,\phi_\ell^i)-\delta(x_n)} 
{p_X(x_n;\lambda,\vec{q}^i_{\setminus\lambda})} \nonumber\\
&=
\begin{cases}
\frac{1}{\lambda} &x_n \neq 0 \\
\frac{-1}{1-\lambda} &x_n = 0
\end{cases},
\label{eq:lamderiv}
\end{align}
which coincides with the BG expression \eqref{lamderivBG}.
The remaining steps also coincide
with those in the BG case, and so the final EM update for $\lambda$, in the case of a GM,\footnote{The arguments in this section reveal that, under signal priors of the form $p_X(x)=(1-\lambda)\delta(x) + \lambda f_X(x)$, where $f_X(\cdot)$ can be arbitrary, the EM update for $\lambda$ is that given in \eqref{lamresult}.}
is given by \eqref{lamresult}.

We next derive the EM updates for the GM parameters $\vec{\omega},\vec{\theta},$ and $\vec{\phi}$.
For each $k = 1,\dots,L$, we incrementally update $\theta_k$, then $\phi_k$, and then the entire vector $\vec{\omega}$, while holding all other parameters fixed.
The EM updates are thus 
\begin{eqnarray}
{\theta}_k^{i+1} 
&=& \argmax_{\theta_k \in \mathbb{R}}\sum_{n=1}^N \hat{\E}\big\{\ln p_X(x_n; \theta_k, \vec{q}^i_{\setminus \theta_k}) \biggiv \vec{y}; \vec{q}^i\big\}, 
	\label{eq:MLtheta}\\
{\phi}_k^{i+1} 
&=& \argmax_{\phi_k > 0}\sum_{n=1}^N \hat{\E}\big\{\ln p_X(x_n; \phi_k, \vec{q}_{\setminus \phi_k}^i) \biggiv \vec{y}; \vec{q}^i\big\} 
	\label{eq:MLphi} \\
{\vec{\omega}}^{i+1} 
&=& \hspace{-3mm}\argmax_{\vec{\omega}>0:\,\sum_k\!\omega_k=1}\sum_{n=1}^N \hat{\E}\big\{\ln p_X(x_n; \vec{\omega}, \vec{q}^i_{\setminus \vec{\omega}}) \biggiv \vec{y}; \vec{q}^i\big\}. \quad
	\label{eq:MLomega} 
\end{eqnarray}

Following \eqref{GAMPtheta}, the maximizing value of $\theta_k$ in \eqref{MLtheta} is again necessarily a value of $\theta_k$ that zeros the derivative, i.e., 
\begin{equation}
\label{eq:thetak}
\sum_{n=1}^{N} \int_{x_n} p_{X|\vec{Y}}(x_n|\vec{y};\vec{q}^i) \frac{d}{d \theta_k}\ln p_X(x_n;\theta_k,\vec{q}_{\setminus \theta_k}^i) = 0,
\end{equation}
Plugging in the derivative 
\small
\begin{align}
&\frac{d}{d\theta_k}\ln p_X(x_n; \theta_k, \vec{q}^i_{\setminus \theta_k}) 
=\Big(\frac{x_n - \theta_k}{\phi_k^i}\Big) \label{eq:thetakder} \\
&\times
\frac{\lambda^i\omega_k^i \mc{N}(x_n;\theta_k,\phi_k^i)}
{(1-\lambda^i)\delta(x_n) + \lambda^i(\omega_k^i \mc{N} (x_n; \theta_k,\phi_k^i) + \sum_{\ell \neq k} \omega_\ell^i \mc{N}(x_n;\theta_\ell^i,\phi_\ell^i))} 
	\nonumber
\end{align}
\normalsize
and the version of $p_{X|\vec{Y}}(x_n|\vec{y}; \vec{q}^i)$ from \eqref{GAMPpost},
integrating \eqref{thetak} separately over $\epsball$ and $\epsballC$ as in \eqref{GAMPlam2BG}, and taking $\epsilon\rightarrow 0$, we find that the $\epsball$ portion vanishes, giving the necessary condition
\begin{equation}
\label{eq:thetak2}
\sum_{n=1}^{N} \int_{x_n}  \!\!\!
\frac{p(x_n|x_n \neq0,\vec{y};\vec{q}^i)\lambda^i \omega_k^i\mc{N}(x_n;\theta_k,\phi_k^i)(x_n - \theta_k)}
{\zeta_n\big(\omega_k^i \mc{N} (x_n; \theta_k,\phi_k^i) + \sum_{\ell \neq k} \omega_\ell^i \mc{N}(x_n;\theta_\ell^i,\phi_\ell^i)\big)}   = 0.
\end{equation}
Since this integral cannot be evaluated in closed form, we apply the approximation $\mc{N}(x_n;\theta_k,\phi_k^i) \approx \mc{N}(x_n;\theta_k^i,\phi_k^i)$ in both the numerator and denominator, and subsequently exploit the fact that
$p(x_n|x_n \neq0,\vec{y};\vec{q}^i)
=\mc{N} (x_n; \hat{r}_n, \mu^r_n) 
\sum_{\ell} \omega_\ell^i \mc{N}(x_n;\theta_\ell^i,\phi_\ell^i)$ 
from \eqref{GAMPpost} to cancel terms, and so obtain the (approximated) necessary condition
\begin{equation}
\label{eq:thetak3}
\sum_{n=1}^N \int_{x_n} \frac{\lambda^i \omega_k^i\mc{N} (x_n; \hat{r}_n, \mu^r_n)
\mc{N}(x_n;\theta_k^i,\phi_k^i)}{\zeta_n}
(x_n - \theta_k) = 0.
\end{equation}
We then simplify \eqref{thetak3} using the Gaussian-pdf multiplication rule, and
set $\theta_k^{i+1}$ equal to the value of $\theta_k$ that satisfies \eqref{thetak3}, which can be found to be
\begin{equation}
\label{eq:thetakfinal}
\theta_k^{i+1} = \frac{\sum_{n=1}^N \pi_n \overline{\beta}_{n,k} \gamma_{n,k}} 
{\sum_{n=1}^N \pi_n \overline{\beta}_{n,k}}
\end{equation}
Note from \eqref{GAMPpost2} that $\pi_n \overline{\beta}_{n,k}$ can be interpreted as the probability that $x_n$ originated from the $k^{th}$ mixture component.

For sparse signals $\vec{x}$, we find that learning the GM means $\{\theta_k\}$ using the above EM procedure yields excellent recovery MSE.
However, for ``heavy-tailed'' signals (i.e., whose pdfs have tails that are not exponentially bounded, such as Student's-t), our experience indicates that the EM-learned values of $\{\theta_k\}$ tend to gravitate towards the outliers in $\{x_n\}_{n=1}^N$, resulting in an overfitting of $p_X(\cdot)$ and thus poor reconstruction MSE.
For such heavy-tailed signals, we find that better reconstruction performance is obtained by fixing the means at zero (i.e., $\theta_k^i\!=\!0~\forall k,i$).
Thus, in the remainder of the paper, we consider two modes of operation: a ``sparse'' mode where $\vec{\theta}$ is learned via the above EM procedure, and a ``heavy-tailed'' mode that fixes $\vec{\theta} = \vec{0}$.

Following \eqref{thetak}, the maximizing value of $\phi_k$ in \eqref{MLphi} is necessarily a value of $\phi_k$ that zeroes the derivative, i.e.,
\begin{equation}
\label{eq:phik}
\sum_{n=1}^{N} \int_{x_n} p_{X|\vec{Y}}(x_n|\vec{y};\vec{q}^i) \frac{d}{d \phi_k}\ln p_X(x_n;\phi_k,\vec{q}_{\setminus \phi_k}^i) = 0.
\end{equation}
As for the derivative in the previous expression, we find
\small
\begin{align}
&\frac{d}{d\phi_k}\ln p_X(x_n;\phi_k, \vec{q}_{\setminus \phi_k}^i) 
 = \frac{1}{2} \left(\frac{|x_n - \theta_k^i|^2}{\phi_k^2}-\frac{1}{\phi_k}\right) 
\label{eq:phikder} \\
&\times
 \frac{\lambda^i\omega_k^i \mc{N}(x_n;\theta_k^i,\phi_k)}
{(1-\lambda^i)\delta(x_n) + \lambda^i(\omega_k^i \mc{N} (x_n; \theta_k^i,\phi_k) + \sum_{\ell \neq k} \omega_\ell^i \mc{N}(x_n;\theta_\ell^i,\phi_\ell^i))} .
\nonumber
\end{align}
\normalsize
Integrating \eqref{phik} separately over $\epsball$ and $\epsballC$, as in \eqref{GAMPlam2BG}, and taking $\epsilon\rightarrow 0$, we find that the $\epsball$ portion vanishes, giving
\small
\begin{equation}
\label{eq:phik2}
\sum_{n=1}^{N} \! \int_{x_n} \!\!\!
\frac{p(x_n|x_n \!\neq\! 0,\vec{y};\vec{q}^i)\lambda^i \omega_k^i\mc{N}(x_n;\theta_k^i,\phi_k)/\zeta_n}
{\omega_k^i \mc{N} (x_n; \theta_k^i,\phi_k) \!+\! \sum_{\ell \neq k} \omega_\ell^i \mc{N}(x_n;\theta_\ell^i,\phi_\ell^i)}
\bigg(\!\frac{|x_n - \theta_k^i|^2}{\phi_k}\!-\!1\!\bigg) 
\end{equation}
\normalsize
Similar to \eqref{thetak2}, this integral is difficult to evaluate, and so we again apply the approximation 
$\mc{N}(x_n;\theta_k^i,\phi_k)\approx \mc{N}(x_n;\theta_k^i,\phi_k^i)$ in the numerator and denominator, after which several terms cancel, yielding the necessary condition
\small
\begin{align}
\label{eq:phik3}
\sum_{n=1}^N \int_{x_n} \frac{\mc{N} (x_n; \hat{r}_n, \mu^r_n)
\lambda^i \omega_k^i \mc{N}(x_n;\theta_k^i,\phi_k^i)}{\zeta_n}
\left(\frac{|x_n-\theta_k^i|^2}{\phi_k} -1\right) = 0.
\end{align}
\normalsize
To find the value of $\phi_k$ satisfying \eqref{phik3}, we expand $|x_n - \theta_k^i|^2 = |x_n|^2 - 2 \real(x_n^* \theta_k^i) + |\theta_k^i|^2$ and apply the Gaussian-pdf multiplication rule, which gives
\begin{equation}
\label{eq:phikfinal}
\phi_k^{i+1} \!\!=\!
\frac{ \sum_{n=1}^N \pi_n\overline{\beta}_{n,k} 
\big(|\theta_k^i-\gamma_{n,k}|^2 \!+\! \nu_{n,k}\big)
}{ \sum_{n=1}^N \pi_n\overline{\beta}_{n,k} } .
\end{equation}

%%%%%%%%%%%%%%%%%%%%%%%%%%%%%%%%%%%%%%%%%%%%%%
Finally, the value of the positive $\vec{\omega}$ maximizing \eqref{MLomega} under the pmf constraint $\sum_{k=1}^L \omega_k = 1$
can be found by solving the unconstrained optimization problem 
$\max_{\vec{\omega},\xi} J(\vec{\omega},\xi)$, where $\xi$ is a Lagrange multiplier and 
\begin{align}
&J(\vec{\omega},\xi) \defn\!
\sum_{n=1}^N \hat{\E}\big\{\!\ln p_X(x_n; \vec{\omega}, \vec{q}^i_{\setminus \vec{\omega}}) \biggiv \vec{y}; \vec{q}^i\big\}
\!-\! \xi\bigg(\sum_{\ell=1}^L \omega_\ell \!-\! 1 \bigg)  \nonumber\\
&= \sum_{n=1}^N\int_{x_n} \!\!p_{X|\vec{Y}}(x_n|\vec{y};\vec{q}^i) \ln p_X(x_n; \vec{\omega}, \vec{q}^i_{\setminus \vec{\omega}})
\!-\! \xi\bigg(\sum_{\ell=1}^L \omega_\ell \!-\! 1 \bigg) .
\end{align}
We start by setting $\frac{d}{d\omega_k} J(\vec{\omega},\xi)=0$,
which yields 
\begin{align}
	\sum_{n=1}^N \int_{x_n} \!\!\frac{p_X(x_n;\vec{q}^i) \mc{N}(x_n;\hat{r}_n,\mu^r_n)}
			{\zeta_n}
		\frac{d}{d\omega_k} \ln p_X(x_n; \vec{\omega}, \vec{q}^i_{\setminus \vec{\omega}}) &= \xi .\! \\
\Leftrightarrow~
\sum_{n=1}^N \int_{x_n} \!\!\frac{p_X(x_n;\vec{q}^i) \mc{N}(x_n;\hat{r}_n,\mu^r_n)} {\zeta_n}
		\frac{\lambda^i \mc{N}(x_n;\theta_k^i,\phi_k^i)}
			{p_X(x_n;\vec{\omega},\vec{q}_{\setminus\vec{\omega}}^i)} &= \xi .\!
\end{align}
Like in \eqref{thetak2} and \eqref{phik2}, the above integral is difficult to evaluate,
and so we approximate $\vec{\omega}\approx \vec{\omega}^i$, which reduces the previous equation to 
\begin{equation}
\xi 
= \sum_{n=1}^N \int_{x_n} \frac{ \lambda^i \mc{N}(x_n;\theta_k^i,\phi_k^i)
				\mc{N}(x_n;\hat{r}_n,\mu^r_n) } {\zeta_n} .
				\label{eq:xi1}
\end{equation}
Multiplying both sides by $\omega_k^i$ for $k=1,\dots,L$, summing over $k$, employing
the fact $1=\sum_k \omega_k^i$, and simplifying, we obtain the equivalent condition 
\begin{align}
\xi 
&= \sum_{n=1}^N \int_{x_n} \frac{ \lambda^i \sum_{k=1}^L \omega_k^i \mc{N}(x_n;\theta_k^i,\phi_k^i)
				\mc{N}(x_n;\hat{r}_n,\mu^r_n) } {\zeta_n}  \\
&= \sum_{n=1}^N \pi_n .
				\label{eq:xi2}
\end{align}
Plugging \eqref{xi2} into \eqref{xi1} and multiplying both sides by $\omega_k$,
the derivative-zeroing value of $\omega_k$ is seen to be
\begin{equation}
\omega_k
\!=\! \frac{ \sum_{n=1}^N \!\int_{x_n} \!\! \lambda^i \omega_k \mc{N}(x_n;\theta_k^i,\phi_k^i)\mc{N}(x_n;\hat{r}_n,\mu^r_n) / \zeta_n }
	{\sum_{n=1}^N \pi_n } ,
				\label{eq:omegak1}
\end{equation}
where, if we use $\omega_k\approx \omega_k^i$ on the right of \eqref{omegak1}, then we obtain 
\begin{equation}
\omega_k^{i+1}
= \frac{\sum_{n=1}^N \pi_n \overline{\beta}_{n,k}}
	{\sum_{n=1}^N \pi_n } .
				\label{eq:omegakfinal}
\end{equation}

Although, for the case of GM priors, approximations were used in the derivation of the EM updates \eqref{thetakfinal}, \eqref{phikfinal}, and \eqref{omegakfinal}, it is interesting to note that, in the case of $L=1$ mixture components, these approximate EM-GM updates coincide with the \emph{exact} EM-BG updates derived in \secref{EMBG}.
In particular, the approximate-EM update of the GM parameter $\theta_1$ in \eqref{thetakfinal} coincides with the exact-EM update of the BG parameter $\theta$ in \eqref{EMtheta3}, the approximate-EM update of the GM parameter $\phi_1$ in \eqref{phikfinal} coincides with the exact-EM update of the BG parameter $\phi$ in \eqref{EMphi3}, and the approximate-EM update of the GM parameter $\omega_1$ in \eqref{omegakfinal} reduces to the fixed value $1$.  
Thus, one can safely use the GM updates above in the BG setting without any loss of optimality.

%%%%%%%%%%%%%%%%%%%%%%%%%%%%%%%%%%%%%%%%%%%%%%%%%%%%%%%%%%%%%%%%%%%%%%%%
\subsection{EM Initialization}	\label{sec:Init}

Since the EM algorithm may converge to a local maximum or at least a saddle point of the likelihood function, proper initialization of the unknown parameters $\vec{q}$ is essential.
Here, we propose initialization strategies for both the ``sparse'' and ``heavy-tailed'' modes of operation, for a given value of $L$.
Regarding the value of $L$, we prescribe a method to learn it in \secref{LearningL}. 
However, the fixed choices $L=3$ for ``sparse'' mode and $L=4$ for ``heavy tailed'' mode usually perform well, as shown in \secref{results}.

For the ``sparse'' mode, we set the initial sparsity rate $\lambda^0$ equal to the theoretical noiseless LASSO PTC, i.e.,
$\lambda^0 = \frac{M}{N}\rho_\text{SE}(\frac{M}{N})$, where \cite{Donoho:PNAS:09}
\begin{eqnarray}
\rho_\textsf{SE}(\textstyle\frac{M}{N})
= \max_{c>0} \displaystyle
	\frac{1-\frac{2N}{M}[(1+c^2)\Phi(-c)-c\phi(c)]}
	{1+c^2-2[(1+c^2)\Phi(-c)-c\phi(c)]} 
					\quad \label{eq:PTC}
\end{eqnarray}
describes the maximum value of $\frac{K}{M}$ supported by LASSO for a given $\frac{M}{N}$, and where
$\Phi(\cdot)$ and $\phi(\cdot)$ denote the cdf and pdf of the 
$\mc{N}(0,1)$ distribution, respectively.
Using the energies $||\vec{y}||_2^2$ and $||\vec{A}||_F^2$ and an assumed value of $\SNR^0$, we initialize the noise and signal variances, respectively, as
\begin{equation}
\psi^0 = \frac{ \norm{\vec{y}}_2^2}{(\SNR^0+1)M}, \
\varphi^0 = \frac{\norm{\vec{y}}_2^2 - M \psi^0}{||\vec{A}||_F^2\lambda^0},
\end{equation}
where, in the absence of (user provided) knowledge about the true $\SNR\defn{\norm{\vec{Ax}}_2^2}/{\norm{\vec{w}}_2^2}$, we suggest $\SNR^0\!=\!100$,
because in our experience this value works well over a wide range of true $\SNR$.
Then, we uniformly space the initial GM means $\vec{\theta}^0$ over $[\frac{-L+1}{2L}, \frac{L-1}{2L}]$, and subsequently fit the mixture weights $\vec{\omega}^0$ and variances $\vec{\phi}^0$ to the uniform pdf supported on $[-0.5,0.5]$ (which can be done offline using the standard approach to EM-fitting of GM parameters, e.g., \cite[p.\ 435]{Bishop:Book:07}).
Finally, we multiply $\vec{\theta}^0$ by $\sqrt{12\varphi^0}$ and $\vec{\phi}^0$ by $12\varphi^0$ to ensure that the resulting signal variance equals $\varphi^0$.

For the ``heavy-tailed'' mode, we initialize $\lambda^0$ and $\psi^0$ as above and set, for $k=1,\dots,L$,
\begin{equation}
\omega_k^0 = \frac{1}{L} ,\ 
\phi_k^0 = \frac{k}{\sqrt{L}} \frac{(\norm{\vec{y}}_2^2 - M \psi^0)}{\norm{\vec{A}}_F^2\lambda^0}, 
\text{~and~} \theta_k^0 = 0.
\end{equation}

%%%%%%%%%%%%%%%%%%%%%%%%%%%%%%%%%%%%%%%%%%%%%%%%%%%%%%%%%%%%%%%%%%%%%
\subsection{EM-GM-AMP Summary and Demonstration}	\label{sec:algorithm}

The fixed-$L$ EM-GM-AMP\footnote{Matlab code at \url{http://www.ece.osu.edu/~schniter/EMturboGAMP}.} algorithm developed in the previous sections is summarized in \tabref{EMGMAMP}.
For EM-BG-AMP (as previously described in \cite{Vila:ASIL:11}), one would simply run EM-GM-AMP with $L=1$.

To demonstrate EM-GM-AMP's ability to learn the underlying signal distribution, \figref{pdf} shows examples of the GM-modeled signal distributions learned by EM-GM-AMP in both ``sparse'' and ``heavy-tailed'' modes.
To create the figure, we first constructed the true signal vector $\vec{x}\in\Real^N$ using $N=2000$ independent draws of the true distribution $p_X(\cdot)$ shown in each of the subplots.  
Then, we constructed measurements $\vec{y}=\vec{Ax}+\vec{w}$ by drawing 
$\vec{A}\in\Real^{M\times N}$ with i.i.d $\mc{N}(0,M^{-1})$ elements and $\vec{w}\in\Real^M$ with i.i.d $\mc{N}(0,\sigma^2)$ elements, with $M=1000$ and $\sigma^2$ chosen to achieve $\SNR=25$ dB.
Finally, we ran EM-GM-AMP according to \tabref{EMGMAMP}, and plotted the GM approximation $p_X(x;\vec{q}^i)$ from \eqref{pX} using the learned pdf parameters 
$\vec{q}^i=[\lambda^i,\vec{\omega}^i,\vec{\theta}^i,\vec{\phi}^i,\psi^i]$.
\Figref{pdf} confirms that EM-GM-AMP is successful in learning a reasonable approximation of the unknown true pdf $p_X(\cdot)$ from the noisy compressed observations $\vec{y}$, in both sparse and heavy-tailed modes.

\putFrag{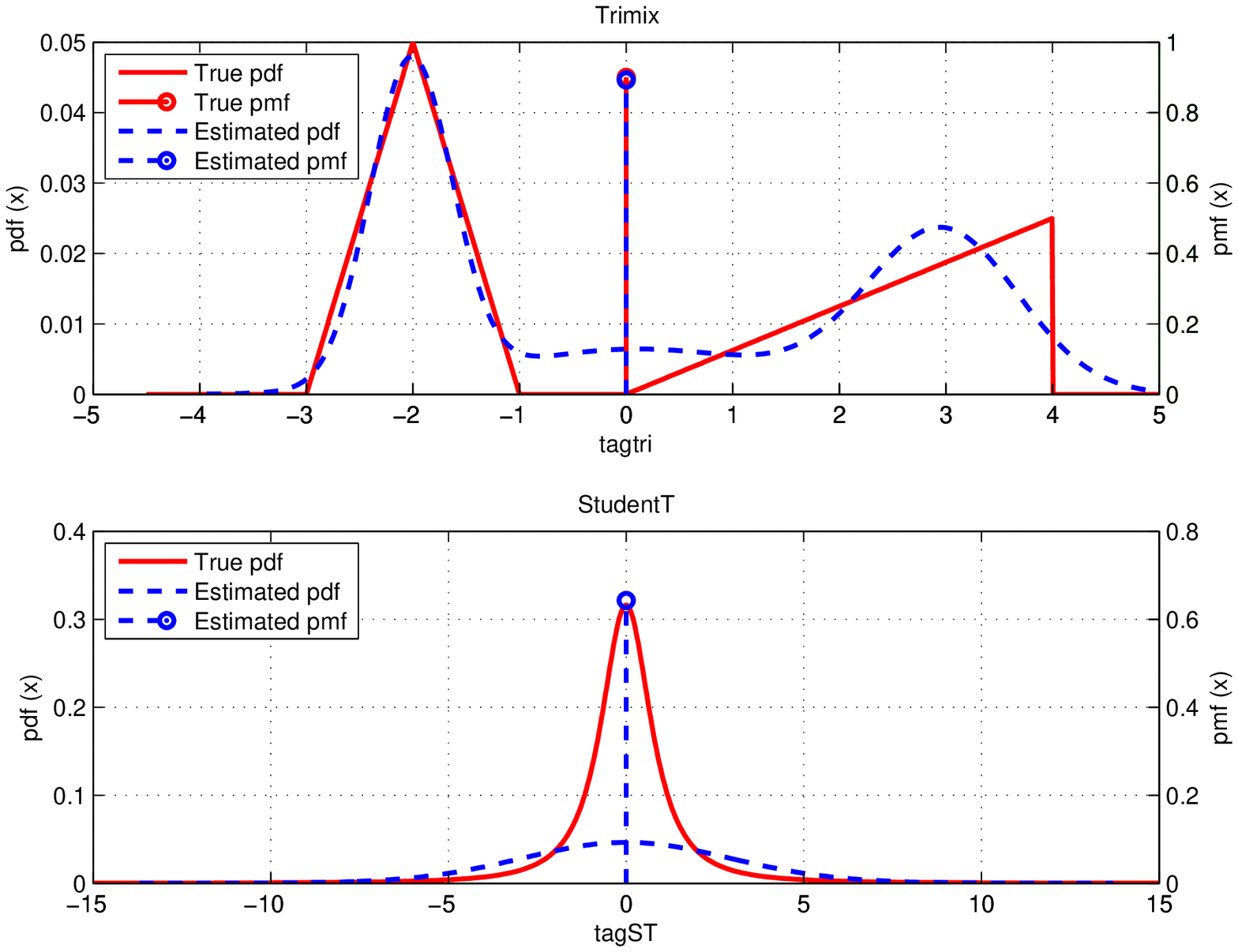}
	{True and EM-GM-AMP-learned versions of the signal distribution 
	 $p_X(x)=\lambda f_X(x) + (1-\lambda)\delta(x)$.  
	 The top subplot shows ``sparse'' mode EM-GM-AMP run using 
	 GM-order $L=3$ on a sparse signal whose non-zero components
	 were generated according to a triangular mixture, whereas the 
	 bottom subplot shows ``heavy-tailed'' EM-GM-AMP run using $L=4$
	 on a Student's-t signal with rate parameter $q=1.67$ 
	 (defined in \eqref{ST}).
	 The density of the continuous component $\lambda f_X(x)$ is marked 
	 on the left axis, while the mass of the discrete component
	 $(1-\lambda)\delta(x)$ is marked on the right axis.}
	{\figsize}
	{\newcommand{\sz}{0.8}
	 \psfrag{pdf (x)}[b][b][\sz]{$\lambda f_X(x)$}
	 \psfrag{pmf (x)}[b][t][\sz][180]{$1-\lambda$}
	 \psfrag{tagtri}[t][t][0.9]{$x$}
	 \psfrag{tagST}[t][t][0.9]{$x$}
	 \psfrag{Trimix}[t][t][0.9]{}
	 \psfrag{StudentT}[t][t][0.9]{}
	}

\putTable{EMGMAMP}
{The EM-GM-AMP algorithm (fixed-$L$ case)}
{
\fbox{
\begin{minipage}{3.3in}
\begin{algorithmic}
\STATE Initialize $L$ and $\vec{q}^0$ as described in \secref{Init}.
\STATE Initialize $\hvec{x}^0=\vec{0}$.
\FOR{ $i=1$ to $I_{\max}$ }
\STATE Generate 
$\hvec{x}^{i}$, 
$\hvec{z}^{i}$, 
$(\vec{\mu}^z)^{i}$, 
$\vec{\pi}^{i}$, 
$\{\vec{\beta}_k^{i},\vec{\gamma}_k^{i},\vec{\nu}_k^{i}\}_{k=1}^L$ 
using GM-GAMP with $\vec{q}^{i-1}$ (see \tabref{gamp}).
\IF{ $\|\hvec{x}^{i} - \hvec{x}^{i-1} \|_2^2 < \tau_{\textsf{em}} \|\hvec{x}^{i-1}\|_2^2$ }
\STATE{break.}
\ENDIF
\STATE Compute $\lambda^{i}$ from $\vec{\pi}^{i-1}$ as described in \eqref{lamresult}.
\FOR{$k=1$ to $L$}
\IF{sparse mode enabled}
\STATE Compute $\theta_k^{i}$ from $\vec{\pi}^{i-1}$, $\vec{\gamma}_k^{i-1}$, $\{\vec{\beta}_l^{i-1}\}_{l=1}^L$ as described in \eqref{thetakfinal}.
\ELSIF{heavy-tailed mode enabled}
\STATE Set $\theta_k^{i}=0$.
\ENDIF
\STATE Compute $\phi_k^{i}$ from $\theta_k^{i-1}$, $\vec{\pi}^{i-1}$, $\vec{\gamma}_k^{i-1}$, $\vec{\nu}_k^{i-1}$, $\{\vec{\beta}_l^{i-1}\}_{l=1}^L$ as described in \eqref{phikfinal}.
\STATE Compute $\vec{\omega}^{i}$ 
from $\vec{\pi}^{i-1}$ and $\{\vec{\beta}_l^{i-1}\}_{l=1}^L$ as described in \eqref{omegakfinal}. 
\ENDFOR
\STATE Compute $\psi^{i}$ from $\hvec{z}^{i}$ and $(\vec{\mu}^z)^{i}$ as in \eqref{EMpsi2}.
\ENDFOR
\end{algorithmic}
\end{minipage}
}
}

%%%%%%%%%%%%%%%%%%%%%%%%%%%%%%%%%%%%%%%%%%%%%%%%%%%%%%%%%%%%%%%%%%%%%%%%
\subsection{Selection of GM Model Order $L$}	\label{sec:LearningL}

We now propose a method to learn the number of GM components, $L$, based on standard maximum likelihood (ML)-based model-order-selection methodology \cite{Stoica:SPM:04}, i.e.,
\begin{align}
\argmax_{L\in \Int^+} ~\ln p(\vec{y};\hvec{q}_L) - \eta(L) ,	\label{eq:MOS}
\end{align}
where $\hvec{q}_L$ is the ML estimate of $\vec{q}$ under the hypothesis $L$ and $\eta(L)$ is a penalty term.
For $\eta(L)$, there are several possibilities, but we focus on the Bayesian information criterion (BIC) \cite{Stoica:SPM:04}: 
\begin{align}
\eta\bic(L)=|\hvec{q}_L|\ln U, 	\label{eq:BIC}
\end{align}
where $|\hvec{q}_L|$ denotes the number\footnote{In our case, the parameters affected by $L$ are the GM means, variances, and weights, so that, for real-valued signals, we use $|\hvec{q}_L|=3L-1$ in ``sparse'' mode and $|\hvec{q}_L|=2L-1$ in heavy-tailed mode, and for complex-valued signals, we use $|\hvec{q}_L|=4L-1$ in ``sparse'' mode and $|\hvec{q}_L|=2L-1$ in heavy-tailed mode.} of real-valued parameters affected by $L$, and 
$U$ is the sample size (see below).

Because $\ln p(\vec{y};\hvec{q}_L)$ is difficult to evaluate, we work with the lower bound (where for now $\Lold$, $\hvec{q}_L$, and $\hvec{q}_{\Lold}$ are arbitrary)
\begin{align}
\lefteqn{
\ln p(\vec{y};\hvec{q}_L)
= \ln \int_{\vec{x}} p(\vec{x}|\vec{y};\hvec{q}_{\Lold})
	\frac{p(\vec{x},\vec{y};\hvec{q}_L)}{p(\vec{x}|\vec{y};\hvec{q}_{\Lold})} 
\label{eq:loglike} }\\
&\geq \int_{\vec{x}} p(\vec{x}|\vec{y};\hvec{q}_{\Lold})
	\ln \frac{p(\vec{x},\vec{y};\hvec{q}_L)}{p(\vec{x}|\vec{y};\hvec{q}_{\Lold})} \label{eq:jensen} \\
&= \int_{\vec{x}} p(\vec{x}|\vec{y};\hvec{q}_{\Lold})
	\ln p(\vec{x},\vec{y};\hvec{q}_L) + \const \\
&= \sum_{n=1}^N \int_{x_n} p(x_n|\vec{y};\hvec{q}_{\Lold})
	\ln p_X(x_n;\hvec{q}_L) + \const 	
						\label{eq:bound1}\\
&= \underbrace{ \sum_{n=1}^N \int_{x_n\neq 0} p(x_n|\vec{y};\hvec{q}_{\Lold})
	\ln f_X(x_n;\hvec{q}_L) }_{
	\displaystyle \defn \LL(\vec{y};\hvec{q}_L)
	} +\, \const , \nonumber\\[-6mm]			\label{eq:LL}
\end{align}
where \eqref{jensen} applies Jensen's inequality, ``\const'' denotes a constant term w.r.t $L$, and \eqref{bound1} holds because $\ln p(\vec{x},\vec{y};\hvec{q}_L) = \ln p(\vec{x};\hvec{q}_L) + \ln p(\vec{y}|\vec{x};\hat{\psi}) = \sum_{n=1}^N \ln p_X(x_n;\hvec{q}_L) + \const$.
Equation \eqref{LL} can then be obtained integrating \eqref{bound1} separately over $\epsball$ and $\epsballC$ and taking $\epsilon\!\rightarrow\! 0$, as done several times in \secref{EMBG}.
Using this lower bound in place of $\ln p(\vec{y};\hvec{q}_L)$ in \eqref{MOS}, we obtain the BIC-inspired model order estimate (where now $\hvec{q}_L$ is specifically the ML estimate of $\vec{q}_L$)
\begin{align}
L^{j+1} 
\defn \argmax_{L\in\Int^+} ~\LL(\vec{y};\hvec{q}_L) - \eta\bic(L) .
						\label{eq:MOS2}
\end{align}%
\vspace{-3mm}

We in fact propose to perform \eqref{MOS2} iteratively, with $j=0,1,2,\dots$ denoting the iteration index.
Notice that \eqref{MOS2} can be interpreted as a ``penalized'' EM update for $L$; if we neglect the penalty term $\eta(L)$, then \eqref{loglike}-\eqref{LL} becomes a standard derivation for the EM-update of $L$ (recall, e.g., the EM derivation in \secref{EMalg}).
The penalty term is essential, though, because the unpenalized log-likelihood lower bound $\LL(\vec{y};\hvec{q}_L)$ is non-decreasing\footnote{Note that $\LL(\vec{y};\hvec{q}_L)$ can be written as a constant plus a scaled value of the negative KL divergence between $p(\vec{x}\giv\vec{x}\!\neq\!\vec{0},\vec{y};\hvec{q}_{\Lold})$ and the GMM $f_X(\vec{x};\hvec{q}_L)$, where the KL divergence is clearly non-increasing in $L$.} in $L$.

We now discuss several practical aspects of our procedure.
First, we are forced to approximate the integral in \eqref{LL}. 
To start, we use GM-GAMP's approximation of the posterior $p(x_n|\vec{y};\hvec{q}_{\Lold})$ from \eqref{GAMPpost}, and the EM approximations of the ML-estimates $\hvec{q}_{\Lold}$ and $\hvec{q}_{L}$ outlined in \Secref{EMGM}.
In this case, the integral in \eqref{LL} takes the form
\begin{align}
\int_{x_n} \!\!\!\!\pi_n \sum_{l=1}^{L^j} \overline{\beta}_{n,l} \mc{N}(x_n;\gamma_{n,l},\nu_{n,l}) \ln \sum_{k=1}^L \omega_k \mc{N}(x_n;\theta_k,\phi_k)  \label{eq:int}
\end{align}
which is still difficult due to the log term.
Hence, we evaluate \eqref{int} using the point-mass approximation $\mc{N}(x_n;\gamma_{n,l},\nu_{n,l})\approx \delta(x_n\!-\!\gamma_{n,l})$.
Second, for the BIC penalty \eqref{BIC}, we use the sample size $U=\sum_{n=1}^N\pi_n$, which is the effective number of terms in the sum in \eqref{LL}.
Third, when maximizing $L$ over $\Int^+$ in \eqref{MOS2}, we start with $L=1$ and increment $L$ in steps of one until the penalized metric decreases.
Fourth, for the initial model order $L^0$, we recommend using $L^0=3$ in ``sparse'' mode and $L^0=4$ in ``heavy-tailed'' mode, i.e., the fixed-$L$ defaults from \secref{Init}.
Finally, \eqref{MOS2} is iterated until either $L^{j+1}=L^{j}$ or a predetermined maximum number of allowed model-order iterations $J_{\max}$ has been reached.

As a demonstration of the proposed model-order selection procedure, we estimated a realization of $\vec{x}$ with $N=1000$ coefficients drawn i.i.d from the triangular mixture pdf shown in \figref{pdf} (top, red) with $\lambda = 0.1$, from the $M=500$ noisy measurements $\vec{y} = \vec{Ax} + \vec{w}$, where $\vec{A}$ was i.i.d $\mc{N}(0,M^{-1})$, and $\vec{w}$ was AWGN such that $\SNR = 20$ dB. 
For illustrative purposes, we set the initial model order at $L^0=1$. 
Iteration $j=1$ yielded the metric $\LL(\vec{y};\hvec{q}_L)-\eta\bic(L)$ shown at the top of \figref{MODSEL}, which was maximized by $L=3\defn L^1$.
The metric resulting from iteration $j=2$ is shown in the middle of \figref{MODSEL}, which was maximized by $L=2\defn L^2$.
At iteration $j=3$, we obtained the metric at the bottom of \figref{MODSEL}, which is also maximized by $L=2\defn L^3$.
Since $L^3=L^2$, the algorithm terminates with final model order estimate $L=2$.
\Figref{MODSEL} also indicates the per-iteration MSE, which is best at the final model order.

\putFrag{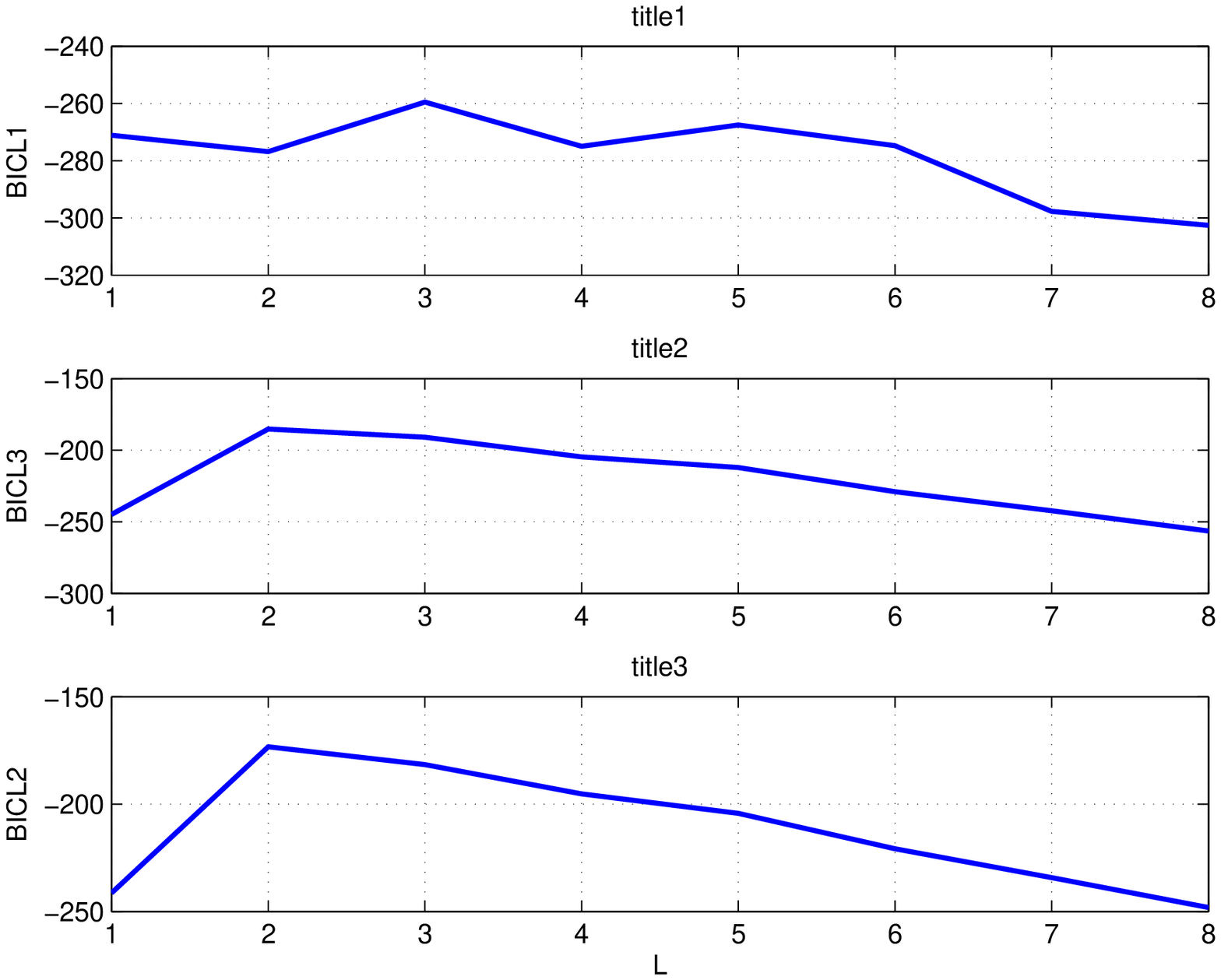}
	{An example of the model-order metric in \eqref{MOS2} over several iterations $j=1,2,3$ using initial model-order $L^j|_{j=0} = 1$, together with the $\NMSE$ of the resulting estimates.}
	{\figsize}
	{\newcommand{\sz}{0.6}
	 \psfrag{BICL1}[b][b][\sz]{$j = 1$}
	 \psfrag{BICL3}[b][b][\sz]{$j = 2$}
	 \psfrag{BICL2}[b][b][\sz]{$j = 3$}
	 \psfrag{title1}[b][b][\sz]{$\NMSE = -29.78$ dB}
	 \psfrag{title2}[b][b][\sz]{$\NMSE = -30.29$ dB}
     \psfrag{title3}[b][b][\sz]{$\NMSE = -30.35$ dB}
	 \psfrag{L}[t][t][0.8]{$L$}
	}

%%%%%%%%%%%%%%%%%%%%%%%%%%%%%%%%%%%%%%%%%%%%%%%%%%%%%%%%%%%%%%%%%%%%%
\section{Numerical Results} 			\label{sec:results}

In this section we report the results of a detailed numerical study that investigate the performance of EM-GM-AMP under both noiseless and noisy settings.
For all experiments, we set the GM-GAMP tolerance to $\tau_{\textsf{gamp}}=10^{-5}$ and the maximum GAMP-iterations to $T_{\max} = 20$ (recall \tabref{gamp}),
and we set the EM tolerance to $\tau_{\textsf{em}}=10^{-5}$ and the maximum EM-iterations to $I_{\max} = 20$ (recall \tabref{EMGMAMP}). 
For fixed-$L$ EM-GM-AMP, we set $L=3$ in ``sparse'' and $L=4$ in ``heavy-tailed'' modes.

%%%%%%%%%%%%%%%%%%%%%%%%%%%%%%%%%%%%%%%%%%%%%%%%%%%%%%%%%%%%%%%%%%%%%
\subsection{Noiseless Phase Transitions}		\label{sec:phase}

We first describe the results of experiments that computed noiseless empirical 
phase transition curves (PTCs) under three sparse-signal distributions.
To evaluate each empirical PTC, we fixed $N=1000$ and constructed a 
$30\times 30$ grid where $(M,K)$ were chosen to yield a uniform sampling of
oversampling ratios $\frac{M}{N}\in[0.05,0.95]$ and
sparsity ratios $\frac{K}{M}\in[0.05,0.95]$.
At each grid point, we generated $R=100$ independent realizations of 
a $K$-sparse signal $\vec{x}$ from a specified distribution
and an $M\times N$ measurement matrix $\vec{A}$
with i.i.d $\mc{N}(0,M^{-1})$ entries.
From the noiseless measurements $\vec{y}=\vec{Ax}$, we 
recovered the signal $\vec{x}$ using several algorithms.
A recovery $\hvec{x}$ from realization $r\in\{1,\dots,R\}$ was defined 
a success if the
$\NMSE \defn \norm{\vec{x}-\hvec{x}}_2^2/\norm{\vec{x}}_2^2 < 10^{-6}$, 
and the average success rate was defined as
$\overline{S} \defn \frac{1}{R}\sum_{r=1}^R S_r$,
where $S_r = 1$ for a success and $S_r=0$ otherwise. 
The empirical PTC was then plotted,
using Matlab's \texttt{contour} command, 
as the $\overline{S}=0.5$ contour over the sparsity-undersampling grid.

\Figsref{BGPT}{BRPT} show the empirical PTCs for five recovery algorithms: 
the proposed EM-GM-AMP algorithm (in ``sparse'' mode) for both $L$ fixed and $L$ learned through model-order selection (MOS),
the proposed EM-BG-AMP algorithm, 
a genie-tuned\footnote{For genie-tuned GM-AMP, for numerical reasons, we set the noise variance at $\psi=10^{-6}$ and, with Bernoulli and BR signals, the mixture variances at $\phi_k=10^{-2}$.} 
GM-AMP that uses the true parameters
$\vec{q}=[\lambda,\vec{\omega},\vec{\theta},\vec{\phi},\psi]$,
and the Donoho/Maleki/Montanari (DMM) LASSO-style AMP from \cite{Donoho:PNAS:09}.
For comparison, 
\figsref{BGPT}{BRPT} also display the theoretical LASSO PTC \eqref{PTC}.  
The signals were generated
as Bernoulli-Gaussian (BG) in \figref{BGPT} (using mean $\theta \!=\! 0$
and variance $\phi \!=\! 1$ for the Gaussian component),
as Bernoulli in \figref{B1PT}
(i.e., all non-zero coefficients set equal to $1$), and
as Bernoulli-Rademacher (BR) in \figref{BRPT}.

For all three signal types, \figsref{BGPT}{BRPT} show 
that the empirical PTC of EM-GM-AMP significantly improves on the 
empirical PTC of DMM-AMP as well as the theoretical PTC of LASSO.
(The latter two are known to converge in the large system limit \cite{Donoho:PNAS:09}.)
For BG signals, \figref{BGPT} shows that
EM-GM-AMP-MOS, EM-GM-AMP, and EM-BG-AMP all yield PTCs that are nearly identical to that of genie-GM-AMP,
suggesting that our EM-learning procedures are working well.
For Bernoulli signals, \figref{B1PT} shows EM-GM-AMP-MOS performing very close to genie-GM-AMP, and both EM-GM-AMP and EM-BG-AMP performing slightly worse but far better than DMM-AMP.
Finally, for BR signals, \figref{BRPT} shows EM-GM-AMP performing significantly better than EM-BG-AMP, since the former is able to accurately model the BR distribution (with $L\geq 2$ mixture components) whereas the latter (with a single mixture component) is not, 
and on par with genie-GM-AMP, whereas  
EM-GM-AMP-MOS performs noticeably better than genie-GM-AMP.
The latter is due to EM-GM-AMP-MOS doing per-realization parameter tuning, while genie-GM-AMP employs the best set of \emph{fixed} parameters over all realizations.

\putFrag{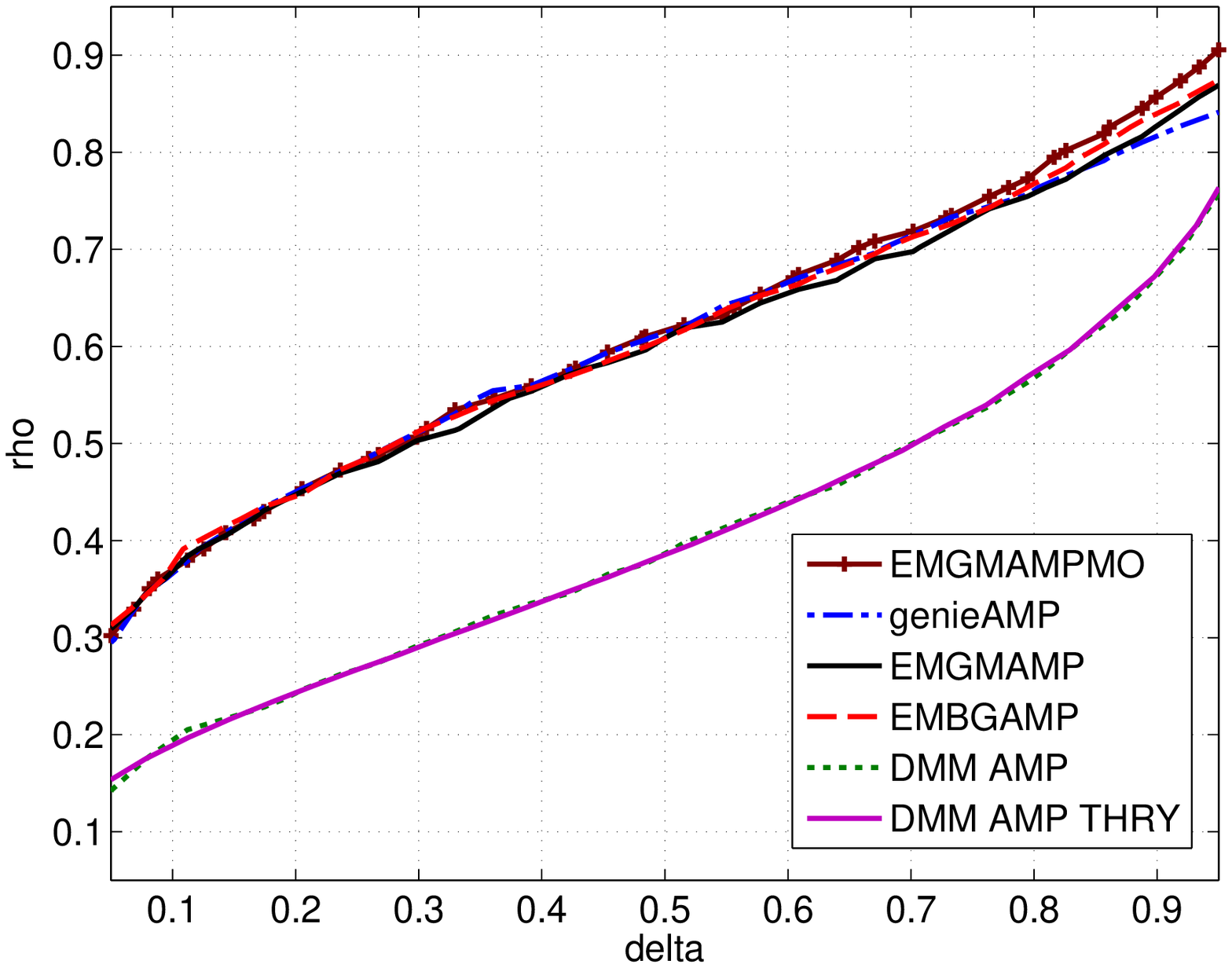}
	{Empirical PTCs and LASSO theoretical PTC 
	for noiseless recovery of Bernoulli-Gaussian signals.}
	{\figsize}
	{\psfrag{delta}[t][t][0.9]{$M/N$} 
	 \psfrag{rho}[][][0.9]{$K/M$} 
         \newcommand{\sz}{0.55}
         \psfrag{EMGMAMPMO}[l][l][\sz]{\sf EM-GM-AMP-MOS}
         \psfrag{EMGMAMP}[l][l][\sz]{\sf EM-GM-AMP}
         \psfrag{EMBGAMP}[l][l][\sz]{\sf EM-BG-AMP}
         \psfrag{genieAMP}[l][l][\sz]{\sf genie GM-AMP}
         \psfrag{DMM AMP}[l][l][\sz]{\sf DMM-AMP}
         \psfrag{DMM AMP THRY}[l][l][\sz]{\sf theoretical LASSO}}
         
\putFrag{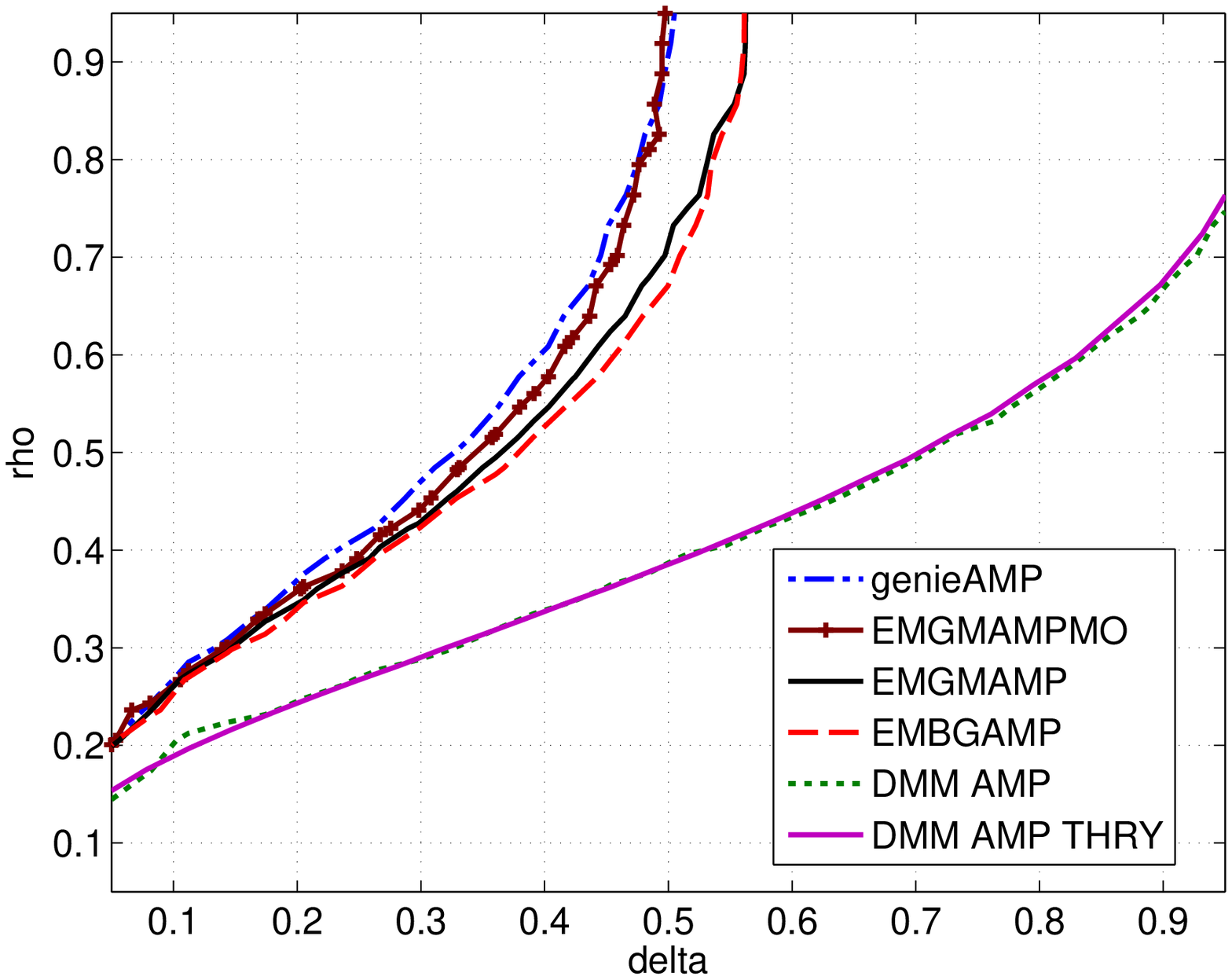}
	{Empirical PTCs and LASSO theoretical PTC 
	for noiseless recovery of Bernoulli signals.}
	{\figsize}
	{\psfrag{delta}[t][t][0.9]{$M/N$} 
	 \psfrag{rho}[][][0.9]{$K/M$} 
         \newcommand{\sz}{0.55}
         \psfrag{EMGMAMPMO}[l][l][\sz]{\sf EM-GM-AMP-MOS}
         \psfrag{EMGMAMP}[l][l][\sz]{\sf EM-GM-AMP}
         \psfrag{EMBGAMP}[l][l][\sz]{\sf EM-BG-AMP}
         \psfrag{genieAMP}[l][l][\sz]{\sf genie GM-AMP}
         \psfrag{DMM AMP}[l][l][\sz]{\sf DMM-AMP}
         \psfrag{DMM AMP THRY}[l][l][\sz]{\sf theoretical LASSO}}

\putFrag{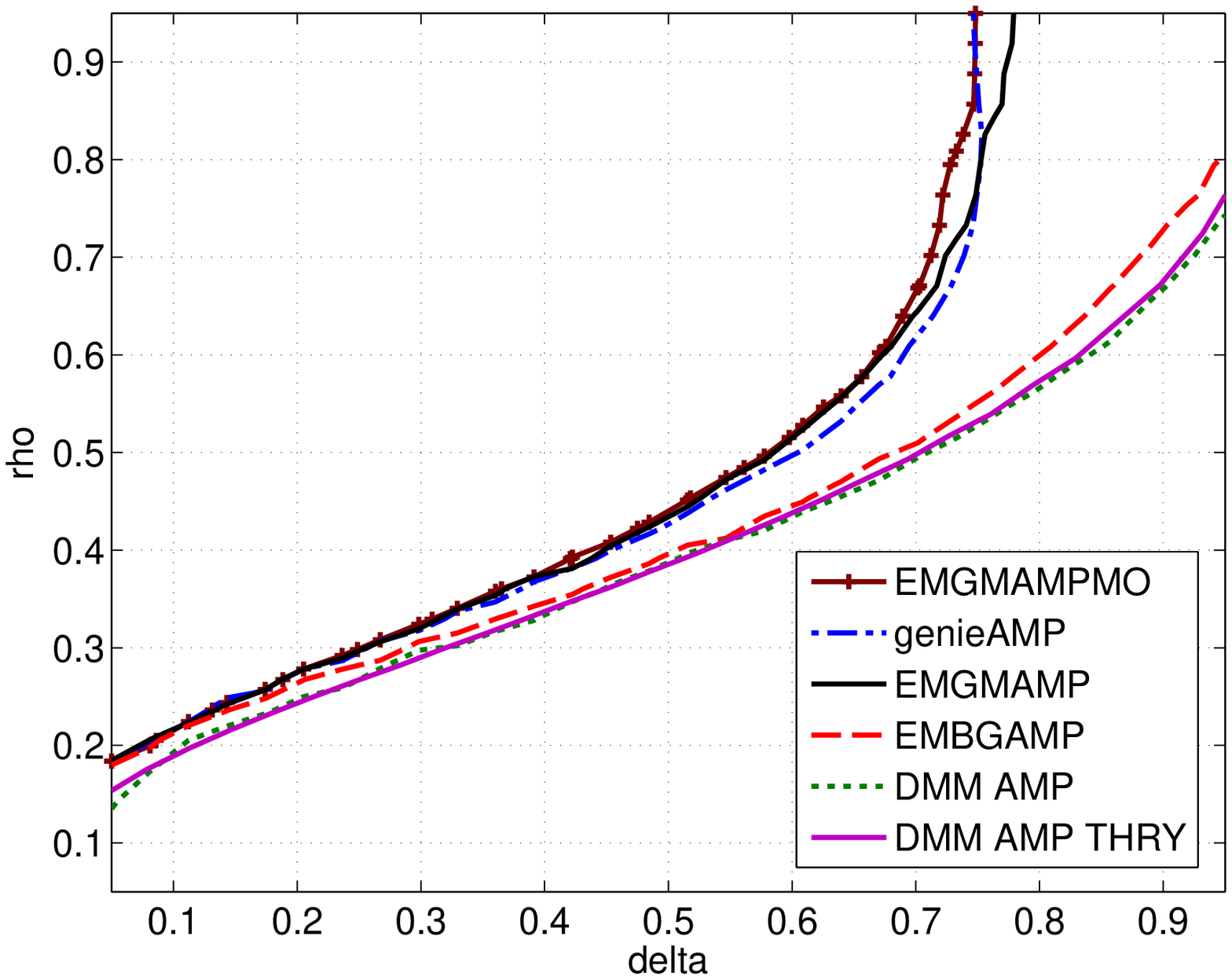}
	{Empirical PTCs and LASSO theoretical PTC 
	for noiseless recovery of Bernoulli-Rademacher signals.}
	{\figsize}
	{\psfrag{delta}[t][t][0.9]{$M/N$} 
	 \psfrag{rho}[][][0.9]{$K/M$} 
         \newcommand{\sz}{0.55}
         \psfrag{EMGMAMPMO}[l][l][\sz]{\sf EM-GM-AMP-MOS}
         \psfrag{EMGMAMP}[l][l][\sz]{\sf EM-GM-AMP}
         \psfrag{EMBGAMP}[l][l][\sz]{\sf EM-BG-AMP}
         \psfrag{genieAMP}[l][l][\sz]{\sf genie GM-AMP}
         \psfrag{DMM AMP}[l][l][\sz]{\sf DMM-AMP}
         \psfrag{DMM AMP THRY}[l][l][\sz]{\sf theoretical LASSO}}
         
To better understand the performance of EM-GM-AMP when $\frac{M}{N}\ll 1$, we fixed $N=8192$ and constructed a $12 \times 9$ grid of $(M,K)$ values spaced uniformly in the log domain.  
At each grid point, we generated $R = 100$ independent realizations of a $K$-sparse BG signal and an i.i.d $\mc{N}(0,M^{-1})$ matrix $\vec{A}$.
We then recovered $\vec{x}$ from the noiseless measurements using EM-GM-AMP-MOS, EM-GM-AMP, EM-BG-AMP, genie-GM-AMP, and the Lasso-solver\footnote{For this experiment, we also tried DMM-AMP but found that it had convergence problems, and we tried SPGL1 but found performance degradations at small $M$.} FISTA\footnote{For FISTA, we used the regularization parameter $\lambda_{\text{\sf FISTA}}=10^{-5}$, which is consistent with the values used for the noiseless experiments in \cite{Beck:JIS:09}.} \cite{Beck:JIS:09}.
\Figref{LOGPTC} shows that the PTCs of EM-GM-AMP-MOS and EM-GM-AMP are nearly identical, slightly better than those of EM-BG-AMP and genie-GM-AMP (especially at very small $M$), and much better than FISTA's.

\putFrag{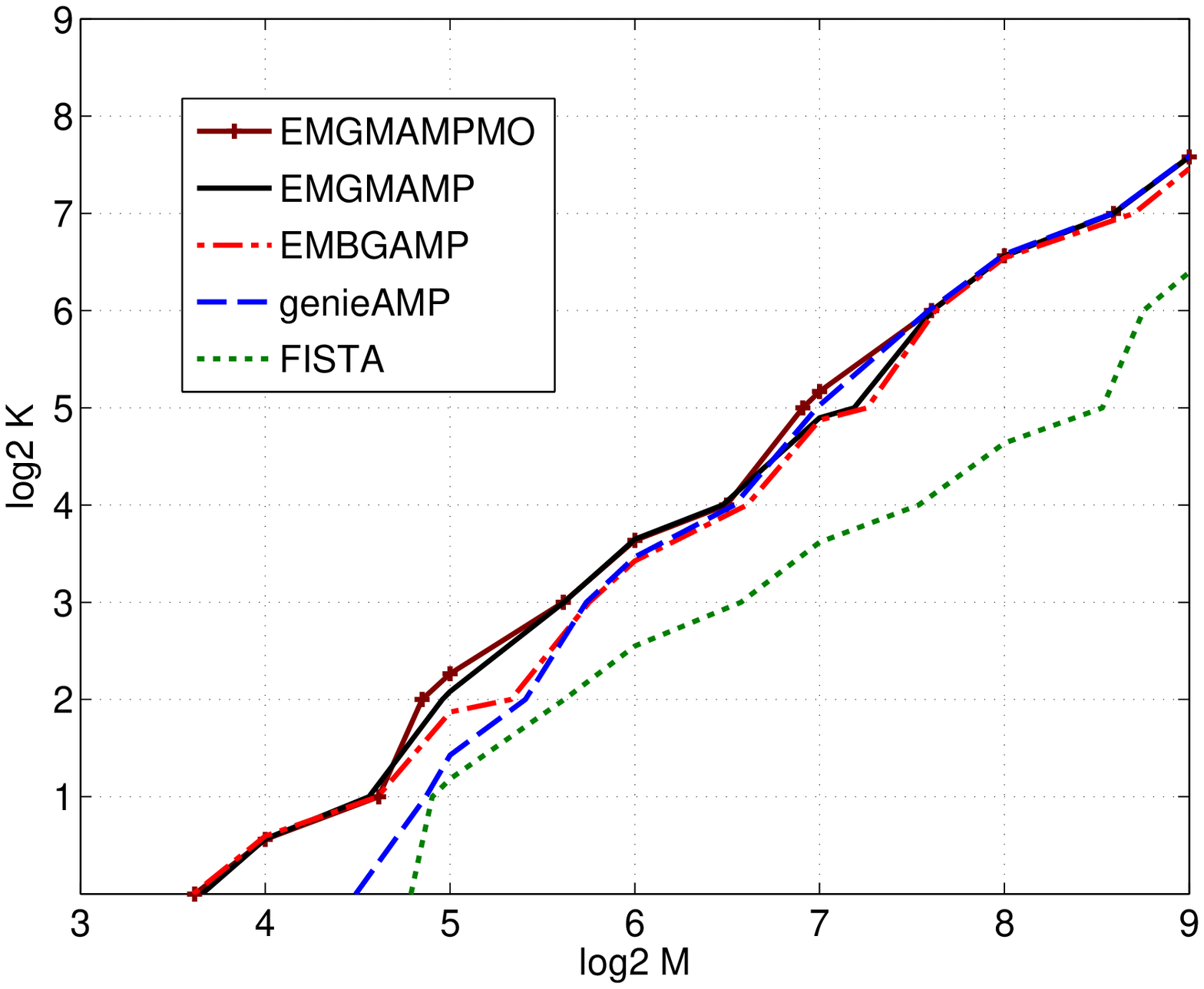}
	{Empirical PTCs for noiseless recovery of Bernoulli-Gaussian signals of length $N=8192$ when $M\ll N$.}
	{\figsize}
	{\psfrag{delta}[t][t][0.9]{$M/N$} 
	 \psfrag{rho}[][][0.9]{$K/M$} 
         \newcommand{\sz}{0.52}
         \psfrag{EMGMAMPMO}[l][l][\sz]{\sf EM-GM-AMP-MOS}
         \psfrag{EMGMAMP}[l][l][\sz]{\sf EM-GM-AMP}
         \psfrag{EMBGAMP}[l][l][\sz]{\sf EM-BG-AMP}
         \psfrag{genieAMP}[l][l][\sz]{\sf genie GM-AMP}
         \psfrag{FISTA}[l][l][\sz]{\sf FISTA}
         \psfrag{log2 M}[l][l][0.85]{\sf $\log_2 M$}
         \psfrag{log2 K}[l][l][0.85]{\sf $\log_2 K$}
         }

Next, we studied the effect of the measurement matrix construction on the performance of EM-GM-AMP in ``sparse'' mode with fixed $L=3$.
For this, we plotted EM-GM-AMP empirical PTCs for noiseless recovery of a length-$N\!=\!1000$ BG signal under several types of measurement matrix $\vec{A}$: 
i.i.d $\mc{N}(0,1)$, 
i.i.d Uniform $[-\tfrac{1}{2},\tfrac{1}{2}]$, 
i.i.d centered Cauchy with scale $1$, 
i.i.d Bernoulli\footnote{For the Bernoulli and BR matrices, we ensured that no two columns of a given realization $\vec{A}$ were identical.} (i.e., $a_{mn}\in\{0,1\}$) with $\lambda_A\defn \Pr\{a_{mn}\neq 0\}=0.15$,
i.i.d zero-mean BR (i.e., $a_{mn}\in\{0,1,-1\}$) with $\lambda_A \in \{0.05,0.15,1\}$, and 
randomly row-sampled Discrete Cosine Transform (DCT).
\Figref{MATCMP} shows that the EM-GM-AMP PTC with i.i.d $\mc{N}(0,1)$ matrices also holds with the other i.i.d zero-mean sub-Gaussian examples (i.e., Uniform and BR with $\lambda_A=1$). 
This is not surprising given that AMP itself has rigorous guarantees for i.i.d zero-mean sub-Gaussian matrices \cite{Bayati:ISIT:12}. 
\Figref{MATCMP} shows that the i.i.d-\mc{N} PTC is also preserved with randomly row-sampled DCT matrices, which is not surprising given AMP's excellent empirical performance with many types of deterministic $\vec{A}$ \cite{Monajemi:PNAS:13} even in the absence of theoretical guarantees.
\Figref{MATCMP} shows, however, that EM-GM-AMP's PTC can degrade with non-zero-mean i.i.d matrices (as in the Bernoulli example) or with super-Gaussian i.i.d matrices (as in the BR example with sparsity rate $\lambda_A=0.05$ and the Cauchy example).
Surprisingly, the i.i.d-$\mc{N}$ PTC is preserved by i.i.d-BR matrices with sparsity rate $\lambda_A = 0.15$, even though $\lambda_A>\frac{1}{3}$ is required for a BR matrix to be sub-Gaussian \cite{Buldygin:Book:00}.

\putFrag{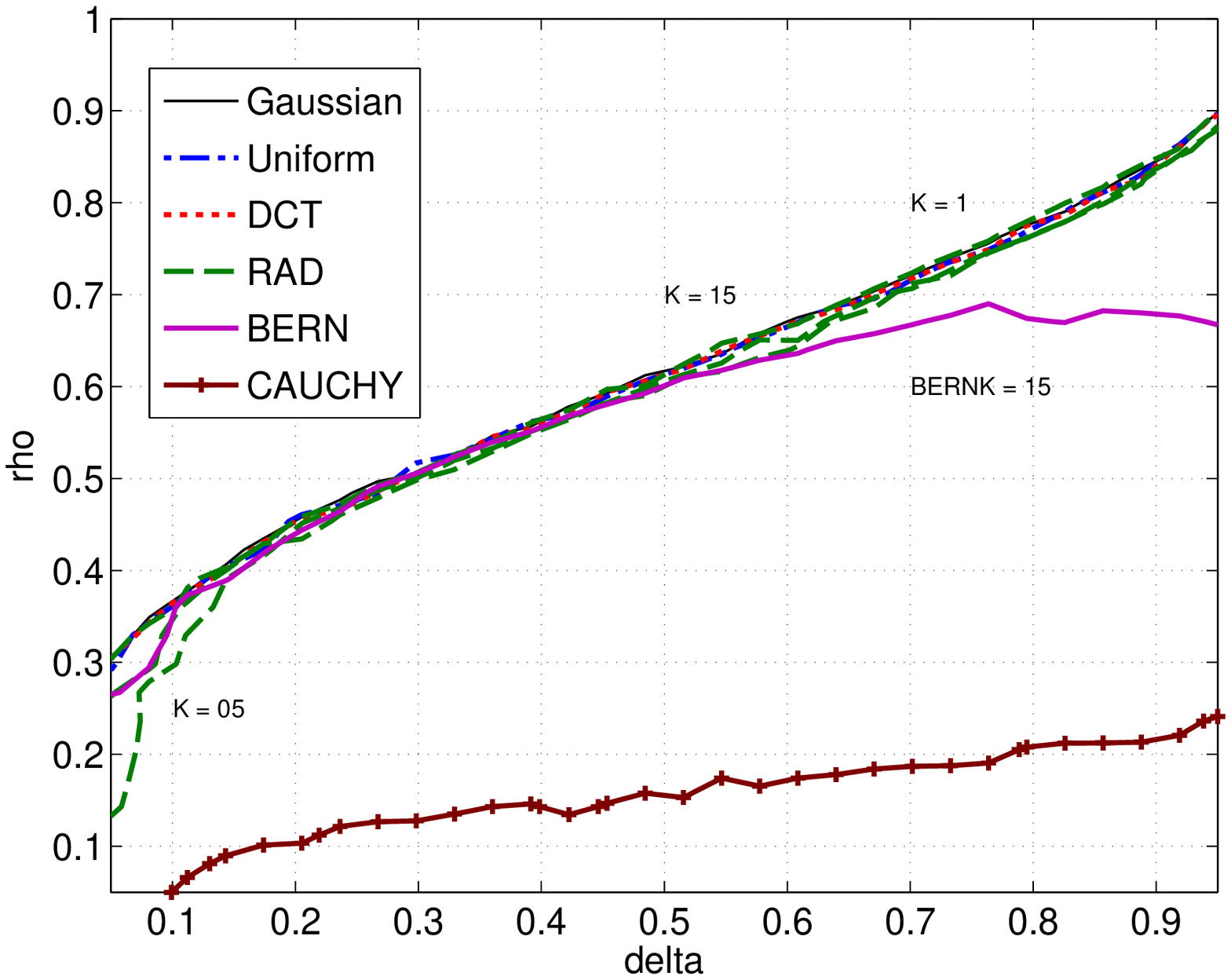}
	{Empirical PTCs for EM-GM-AMP noiseless recovery of Bernoulli-Gaussian signals under various $\vec{A}$: i.i.d $\mc{N}(0,1)$, i.i.d Uniform $[-\tfrac{1}{2},\tfrac{1}{2}]$, i.i.d Bernoulli with $\lambda_A\defn \Pr\{a_{mn}\neq 0\}=0.15$, i.i.d zero-mean Bernoulli-Rademacher with $\lambda_A \in \{0.05,0.15,1\}$, i.i.d Cauchy, and randomly row-sampled DCT.}
	{\figsize}
	{\psfrag{delta}[t][t][0.9]{$M/N$} 
	 \psfrag{rho}[][][0.9]{$K/M$} 
         \newcommand{\sz}{0.55}
         \psfrag{Gaussian}[l][l][\sz]{\sf i.i.d $\mc{N}$}
         \psfrag{Uniform}[l][l][\sz]{\sf i.i.d Unif}
         \psfrag{CAUCHY}[l][l][\sz]{\sf i.i.d Cauchy}
         \psfrag{RAD}[l][l][\sz]{\sf i.i.d BR}
         \psfrag{BERN}[l][l][\sz]{\sf i.i.d Bern}
         \psfrag{DCT}[l][l][\sz]{\sf DCT}
         \psfrag{K = 1}[r][r][0.85]{\sf $\lambda_A=1$}
         \psfrag{K = 05}[l][l][0.85]{\sf $\lambda_A=0.05$}
         \psfrag{K = 15}[r][r][0.85]{\sf $\lambda_A=0.15$}
         \psfrag{BERNK = 15}[r][r][0.6]{}
         }

%%%%%%%%%%%%%%%%%%%%%%%%%%%%%%%%%%%%%%%%%%%%%%%%%%%%%%%%%%%%%%%%%%%%%
\subsection{Noisy Sparse Signal Recovery} 		\label{sec:sigrecovnoise}

\Figsref{BGNMSE}{BRNMSE} show $\NMSE$ for noisy recovery of BG, Bernoulli, and BR signals, respectively.
To construct these plots, we fixed $N=1000$, $K=100$, $\SNR=25$ dB, and varied $M$.
Each data point represents $\NMSE$ averaged over $R=500$ realizations, where in each realization we drew an $\vec{A}$ with i.i.d $\mc{N}(0,M^{-1})$ elements, an AWGN noise vector, and a random signal vector.
For comparison, we show the performance of
the proposed EM-GM-AMP (in ``sparse'' mode) for both MOS and $L=3$ versions,
EM-BG-AMP, 
genie-tuned\footnote{We ran both OMP (using the implementation from \url{http://sparselab.stanford.edu/OptimalTuning/code.htm}) and SP under $10$ different sparsity assumptions, spaced uniformly from $1$ to $2K$, and reported the lowest $\NMSE$ among the results.\label{greedy}}
Orthogonal Matching Pursuit (OMP) \cite{PRK93},
genie-tuned$^{\text{\ref{greedy}}}$ 
Subspace Pursuit (SP) \cite{Dai:TIT:09}, 
Bayesian Compressive Sensing (BCS) \cite{Ji:TSP:08}, 
Sparse Bayesian Learning \cite{Wipf:TSP:04} 
(via the more robust T-MSBL \cite{Zhang:JSTSP:11}), 
de-biased genie-tuned\footnote{\label{spl}We ran SPGL1 in `BPDN' mode:
  $\min_{\hvec{x}}\norm{\vec{x}}_1~\text{s.t.}~\norm{\vec{y}-\vec{Ax}}_2\leq\sigma$, for hypothesized tolerances $\sigma^2\in\{0.1,0.2,\dots,1.5\}\times M\psi$,
  and reported the lowest $\NMSE$ among the results.} 
LASSO (via SPGL1 \cite{vandenBerg:JSC:08}), and 
Smoothed-$\ell_0$ (SL0) \cite{Mohimani:09}.
All algorithms were run under the suggested defaults, 
with \texttt{noise=small} in T-MSBL.

For BG signals, \figref{BGNMSE} shows that EM-GM-AMP-MOS, EM-GM-AMP, and EM-BG-AMP together exhibit the best performance among the tested algorithms,
reducing the $M/N$ breakpoint (i.e., the location of the knee in the $\NMSE$ curve, which represents a sort of phase transition) from $0.3$ down to $0.26$, but also improving $\NMSE$ by $\approx 1$ dB relative to the next best algorithm, which was BCS.
Relative to the other EM-AMP variants, MOS resulted in a slight degradation of performance for $\frac{M}{N}$ between $0.26$ and $0.31$, but was otherwise identical.
For Bernoulli signals, \figref{B1NMSE} shows much more significant gains for EM-GM-AMP-MOS, EM-GM-AMP and EM-BG-AMP over the other algorithms:
the $M/N$ breakpoint was reduced from $0.4$ down to $0.32$ (and even $0.3$ with MOS), and the $\NMSE$ was reduced by $\approx 8$ dB relative to the next best algorithm, which was T-MSBL in this case.
Finally, for BR signals, \figref{BRNMSE} shows a distinct advantage for EM-GM-AMP and EM-GM-AMP-MOS over the other algorithms, including EM-BG-AMP, due to the formers' ability to accurately model the BR signal prior.
In particular, for $M/N\geq 0.36$, EM-GM-AMP-MOS reduces the $\NMSE$ by $10$ dB relative to the best of the other algorithms (which was either EM-BG-AMP or T-MSBL depending on the value of $M/N$) and reduces the $M/N$ breakpoint from $0.38$ down to $0.35$.
 
\putFrag{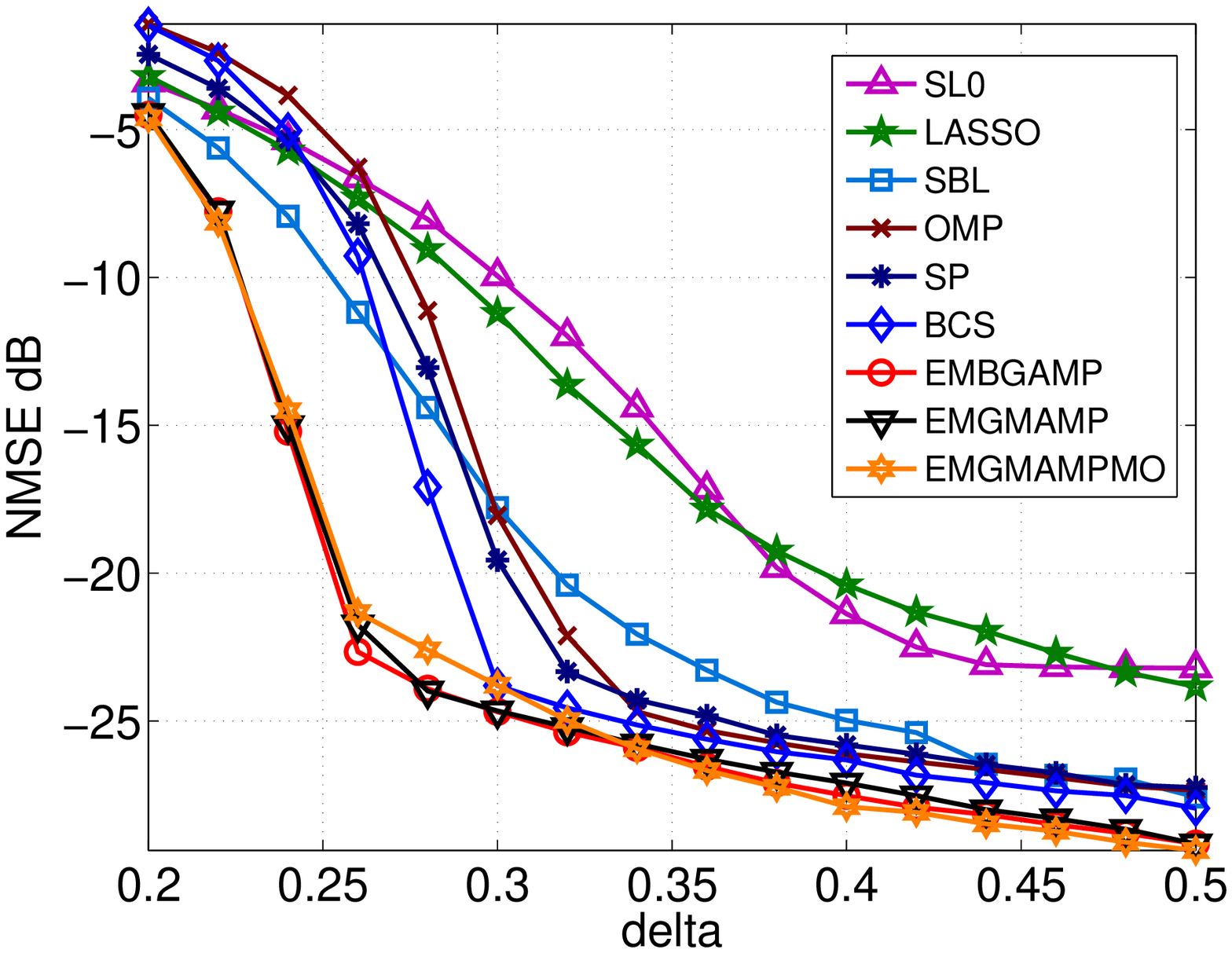}
	{$\NMSE$ versus undersampling ratio $M/N$ for noisy recovery of Bernoulli-Gaussian signals.}
	{\figsize}
	{\psfrag{delta}[t][t][0.9]{$M/N$} 
	 \psfrag{NMSE dB}[][][0.9]{\sf NMSE [dB]}
         \newcommand{\sz}{0.45}
         \psfrag{EMGMAMPMO}[l][l][\sz]{\sf EM-GM-AMP-MOS}
         \psfrag{EMGMAMP}[l][l][\sz]{\sf EM-GM-AMP}
         \psfrag{EMBGAMP}[l][l][\sz]{\sf EM-BG-AMP}
         \psfrag{BCS}[l][l][\sz]{\sf BCS}
         \psfrag{SBL}[l][l][\sz]{\sf T-MSBL}
         \psfrag{SL0}[l][l][\sz]{\sf SL0}
          \psfrag{SP}[l][l][\sz]{\sf genie SP}
         \psfrag{OMP}[l][l][\sz]{\sf genie OMP}
         \psfrag{LASSO}[l][l][\sz]{\sf genie SPGL1}}
         
\putFrag{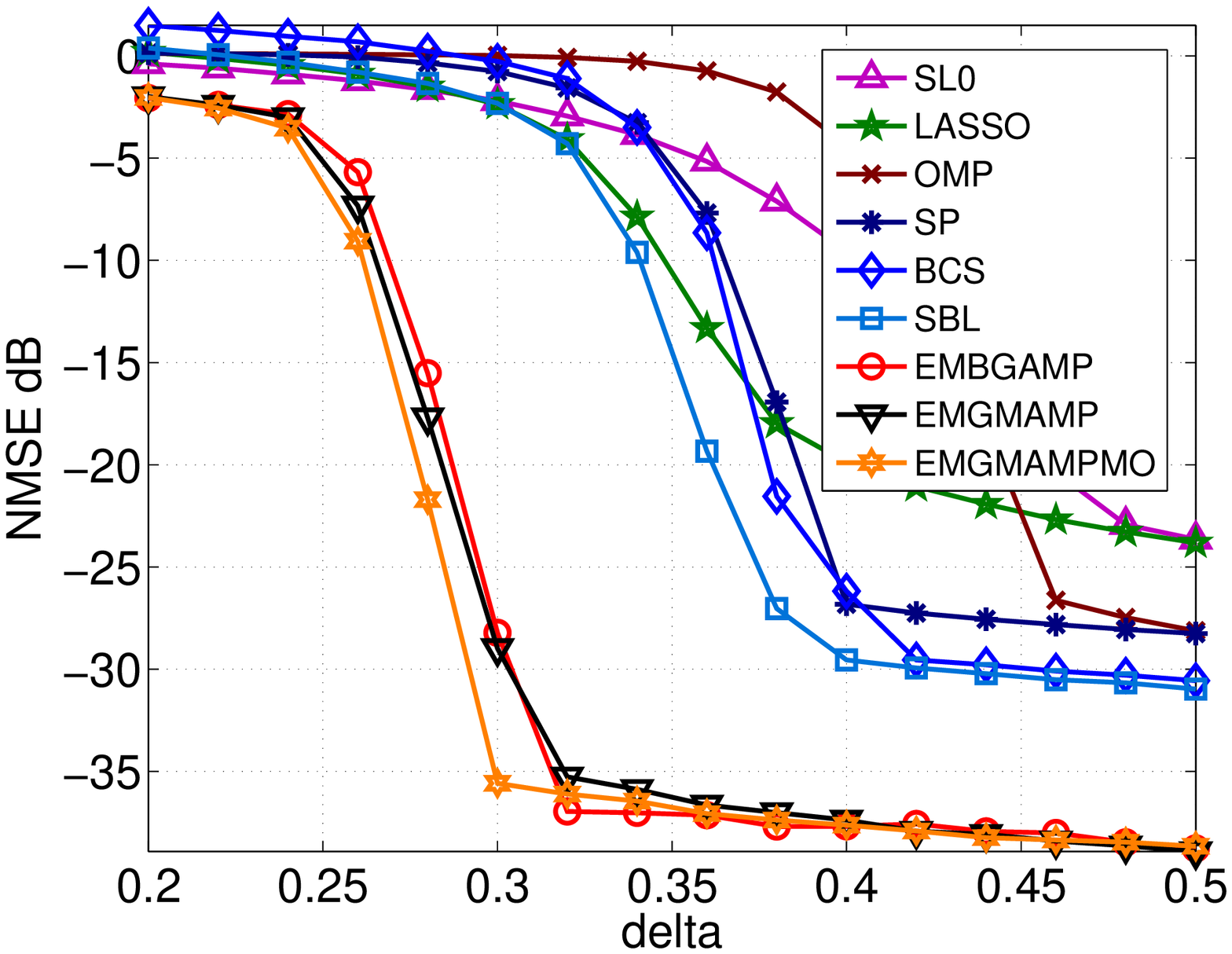}
	{$\NMSE$ versus undersampling ratio $M/N$ for noisy recovery of Bernoulli signals.}
	{\figsize}
	{\psfrag{delta}[t][t][0.9]{$M/N$} 
	 \psfrag{NMSE dB}[][][0.9]{\sf NMSE [dB]}
         \newcommand{\sz}{0.45}
         \psfrag{EMGMAMPMO}[l][l][\sz]{\sf EM-GM-AMP-MOS}
         \psfrag{EMGMAMP}[l][l][\sz]{\sf EM-GM-AMP}
         \psfrag{EMBGAMP}[l][l][\sz]{\sf EM-BG-AMP}
         \psfrag{BCS}[l][l][\sz]{\sf BCS}
         \psfrag{SBL}[l][l][\sz]{\sf T-MSBL}
         \psfrag{SL0}[l][l][\sz]{\sf SL0}
         \psfrag{SP}[l][l][\sz]{\sf genie SP}
         \psfrag{OMP}[l][l][\sz]{\sf genie OMP}
         \psfrag{LASSO}[l][l][\sz]{\sf genie SPGL1}}
         
\putFrag{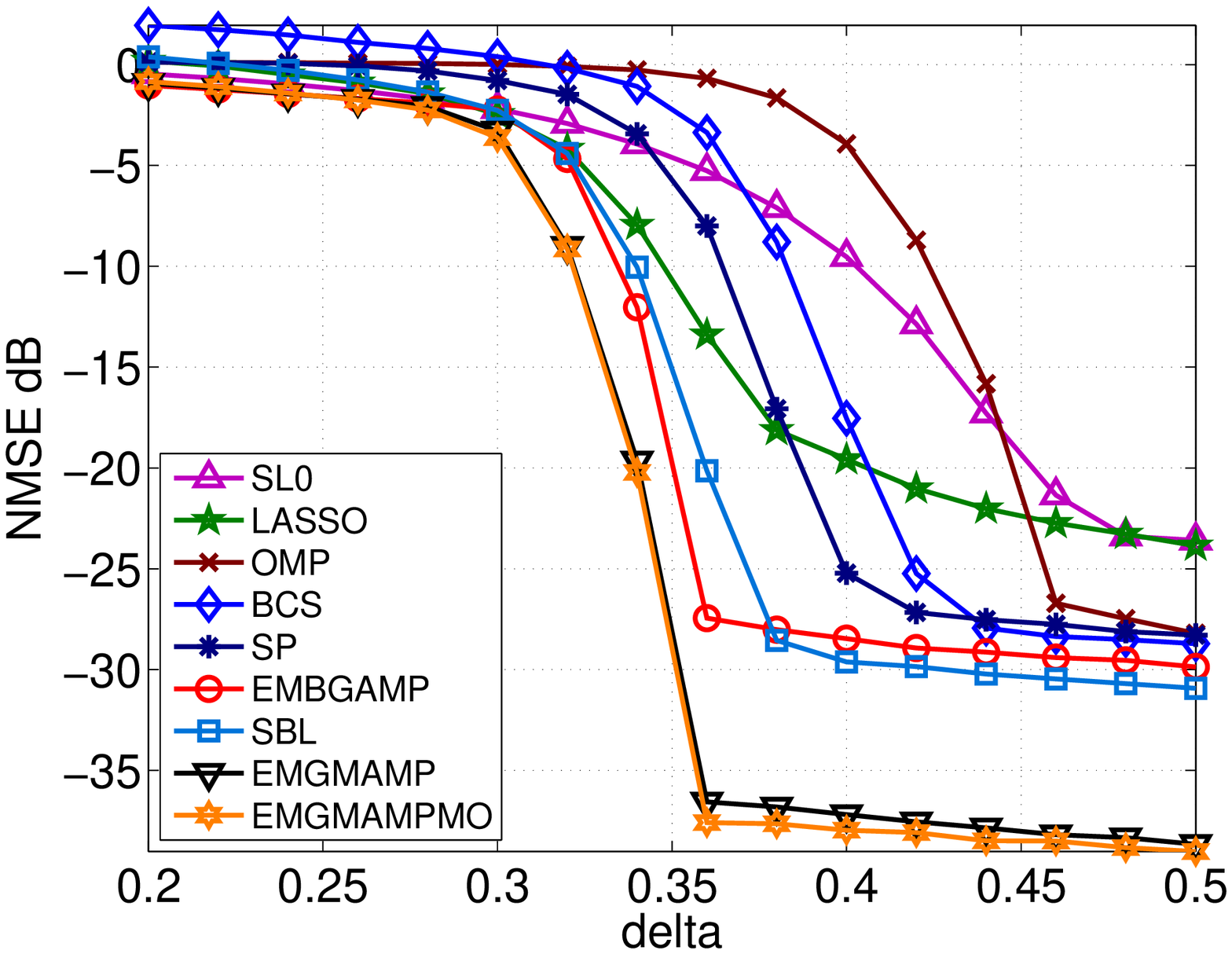}
	{$\NMSE$ versus undersampling ratio $M/N$ for noisy recovery of Bernoulli-Rademacher signals.}
	{\figsize}
	{\psfrag{delta}[t][t][0.9]{$M/N$} 
	 \psfrag{NMSE dB}[][][0.9]{\sf NMSE [dB]}
         \newcommand{\sz}{0.45}
         \psfrag{EMGMAMPMO}[l][l][\sz]{\sf EM-GM-AMP-MOS}
         \psfrag{EMGMAMP}[l][l][\sz]{\sf EM-GM-AMP}
         \psfrag{EMBGAMP}[l][l][\sz]{\sf EM-BG-AMP}
         \psfrag{BCS}[l][l][\sz]{\sf BCS}
         \psfrag{SBL}[l][l][\sz]{\sf T-MSBL}
         \psfrag{SL0}[l][l][\sz]{\sf SL0}
          \psfrag{SP}[l][l][\sz]{\sf genie SP}
         \psfrag{OMP}[l][l][\sz]{\sf genie OMP}
         \psfrag{LASSO}[l][l][\sz]{\sf genie SPGL1}}
         
To investigate each algorithm's robustness to AWGN, we plotted
the $\NMSE$ attained in the recovery of BR signals with $N=1000$, $M = 500$, 
and $K = 100$ as a function of $\SNR$ in \figref{SNRNMSE}, where
each point represents an average over $R=100$ problem realizations, where in each realization we drew an $\vec{A}$ with i.i.d $\mc{N}(0,M^{-1})$ elements, an AWGN noise vector, and a random signal vector.
All algorithms were under the same conditions as those reported previously,
except that T-MSBL used \texttt{noise=small} when $\SNR > 22$dB and \texttt{noise=mild} when $\SNR \leq 22$ dB, as recommended in \cite{ZhilinZhang:11}.
From \figref{SNRNMSE}, we see that the essential behavior observed in the fixed-$\SNR$ BR plot \figref{BRNMSE} holds over a wide range of $\SNR$s. 
In particular, \figref{SNRNMSE} shows that EM-GM-AMP and EM-GM-AMP-MOS yield significantly lower $\NMSE$ than all other algorithms over the full $\SNR$ range, while EM-BG-AMP and T-MSBL yield the second lowest $\NMSE$ (also matched by BCS for $\SNR$s between $30$ and $40$ dB).
Note, however, than T-MSBL must be given some knowledge about the true noise variance in order to perform well \cite{ZhilinZhang:11}, unlike the proposed algorithms.

\putFrag{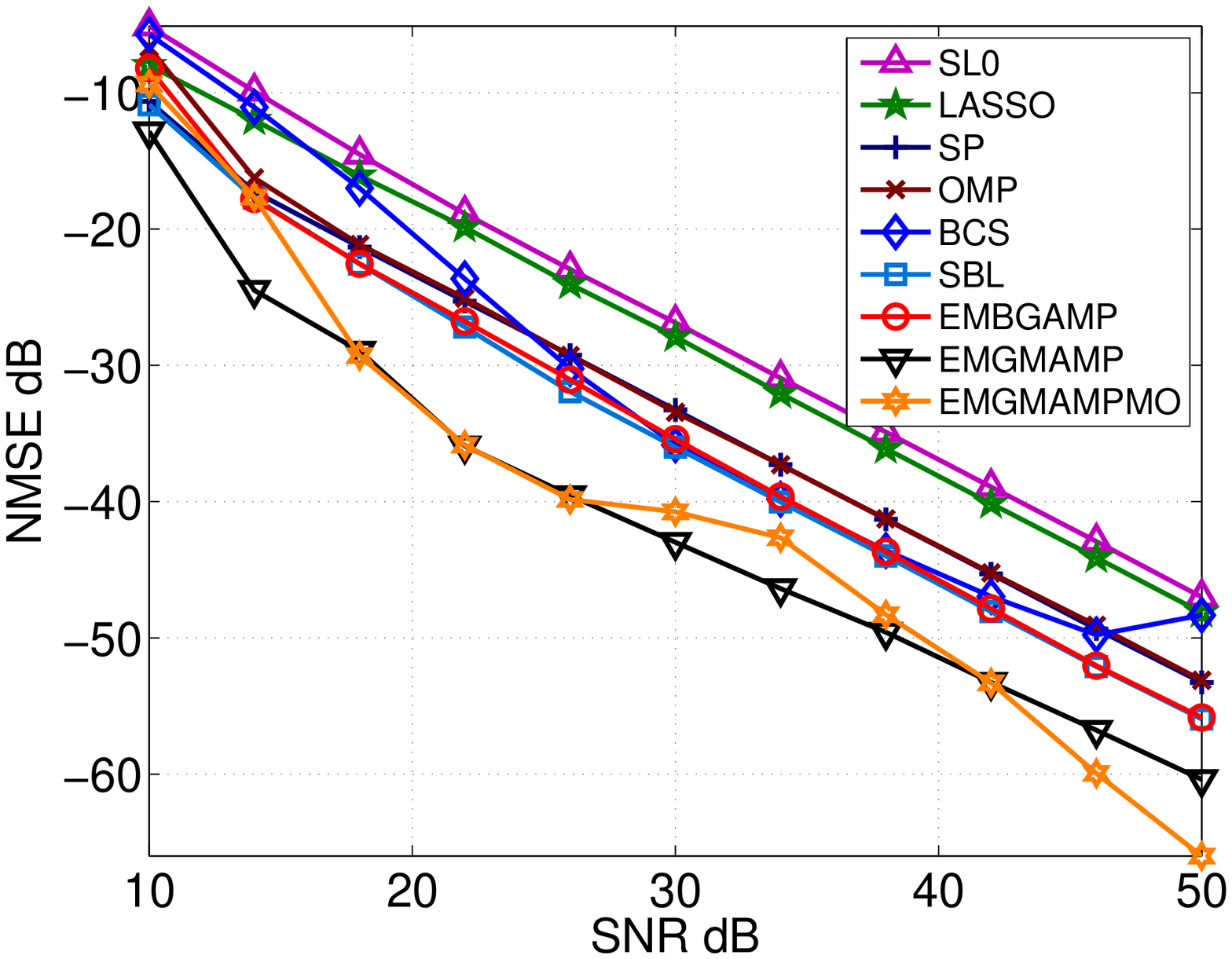}
	{$\NMSE$ versus $\SNR$ for noisy recovery of Bernoulli-Rademacher signals.}
	{\figsize}
	{\psfrag{SNR dB}[t][t][0.9]{\sf SNR [dB]} 
	 \psfrag{NMSE dB}[][][0.9]{\sf NMSE [dB]}
         \newcommand{\sz}{0.4}
         \psfrag{EMGMAMPMO}[l][l][\sz]{\sf EM-GM-AMP-MOS}
         \psfrag{EMGMAMP}[l][l][\sz]{\sf EM-GM-AMP}
         \psfrag{EMBGAMP}[l][l][\sz]{\sf EM-BG-AMP}
         \psfrag{BCS}[l][l][\sz]{\sf BCS}
         \psfrag{SBL}[l][l][\sz]{\sf T-MSBL}
         \psfrag{SL0}[l][l][\sz]{\sf SL0}
         \psfrag{SP}[l][l][\sz]{\sf genie SP}
         \psfrag{OMP}[l][l][\sz]{\sf genie OMP}
         \psfrag{LASSO}[l][l][\sz]{\sf genie SPGL1}}

%%%%%%%%%%%%%%%%%%%%%%%%%%%%%%%%%%%%%%%%%%%%%%%%%%%%%%%%%%%%%%%%%
\subsection{Heavy-Tailed Signal Recovery}		\label{sec:heavy}

In many applications of compressive sensing, the signal to be recovered is not perfectly sparse, but instead contains a few large coefficients and many small ones.  
While the literature often refers to such signals as ``compressible,'' there are many real-world signals that do not satisfy the technical definition of compressibility (see, e.g., \cite{Cevher:NIPS:09}), and so we refer to such signals more generally as ``heavy tailed.''

To investigate algorithm performance for these signals, we first consider an i.i.d Student's-t signal, with prior pdf 
\begin{equation}
\textstyle
p_X(x;q) \defn \frac{\Gamma((q+1)/2))}{\sqrt{\pi} \Gamma(q/2)}\left(1 + x^2\right)^{-(q+1)/2}
\label{eq:ST}
\end{equation}
under the (non-compressible) rate $q = 1.67$, which has been shown to be an excellent model for wavelet coefficients of natural images \cite{Cevher:NIPS:09}. 
For such signals, \figref{STNMSE} plots $\NMSE$ versus the number of measurements $M$ for fixed $N=1000$, $\SNR=25$ dB, and an average of $R=500$ realizations, where in each realization we drew an $\vec{A}$ with i.i.d $\mc{N}(0,M^{-1})$ elements, an AWGN noise vector, and a random signal vector.
\Figref{STNMSE} shows both variants of EM-GM-AMP (here run in ``heavy-tailed'' mode) outperforming all other algorithms under test.\footnote{%
In this experiment, we ran both OMP and SP under $10$ different sparsity hypotheses, spaced uniformly from $1$ to $K\lasso = M \rho_\textsf{SE}(\frac{M}{N})$, and reported the lowest $\NMSE$ among the results.} 
We have also verified (in experiments not shown here) that ``heavy-tailed'' EM-GM-AMP exhibits similarly good performance with other values of the Student's-t rate parameter $q$, as well as for i.i.d centered Cauchy signals.

\putFrag{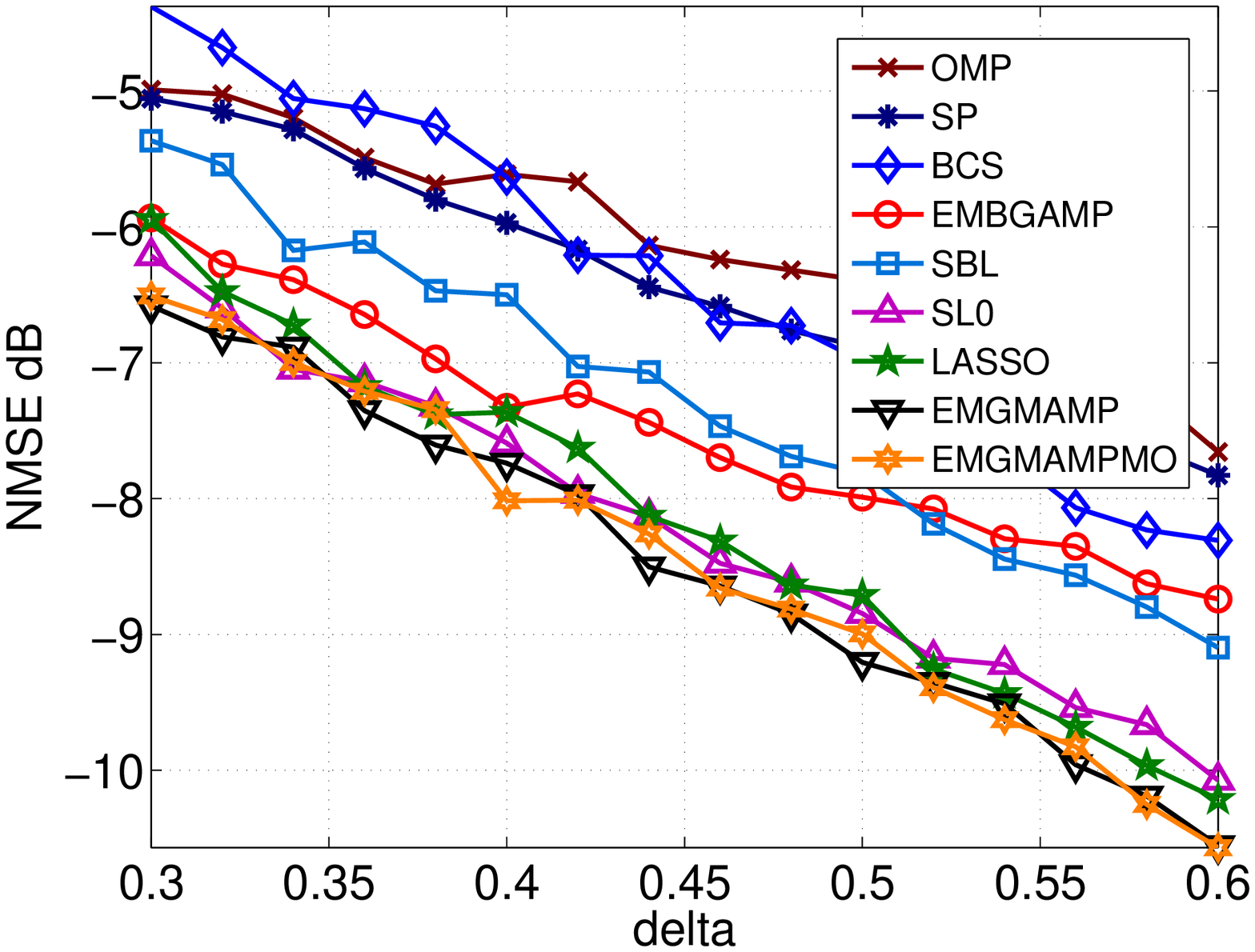}
	{$\NMSE$ versus undersampling ratio $M/N$ for noisy recovery of Student-t signals with rate parameter $1.67$.}
	{\figsize}
	{\psfrag{delta}[t][t][0.9]{$M/N$} 
	 \psfrag{NMSE dB}[][][0.9]{\sf NMSE [dB]}
         \newcommand{\sz}{0.45}
         \psfrag{EMGMAMPMO}[l][l][\sz]{\sf EM-GM-AMP-MOS}
         \psfrag{EMGMAMP}[l][l][\sz]{\sf EM-GM-AMP}
         \psfrag{EMBGAMP}[l][l][\sz]{\sf EM-BG-AMP}
         \psfrag{BCS}[l][l][\sz]{\sf BCS}
         \psfrag{SBL}[l][l][\sz]{\sf T-MSBL}
         \psfrag{SL0}[l][l][\sz]{\sf SL0}
          \psfrag{SP}[l][l][\sz]{\sf genie SP}
         \psfrag{OMP}[l][l][\sz]{\sf genie OMP}
         \psfrag{LASSO}[l][l][\sz]{\sf genie SPGL1}}
 
To investigate the performance for positive heavy-tailed signals,
we conducted a similar experiment using i.i.d log-normal $\vec{x}$, 
generated using the distribution
\begin{equation}
p_X(x;\mu,\sigma^2) = \tfrac{1}{x\sqrt{2\pi \sigma^2}} \exp{-\tfrac{(\ln x - \mu)}{2\sigma^2} }
\end{equation}
with location parameter $\mu = 0$ and scale parameter $\sigma^2 = 1$.
\Figref{LOGNNMSE} confirms the excellent performance of EM-GM-AMP-MOS, EM-GM-AMP, and EM-BG-AMP over all tested undersampling ratios $M/N$.
We postulate that, for signals known apriori to be positive, EM-GM-AMP's performance could be further improved through the use of a prior $p_X$ with support restricted to the the positive reals, via a mixture of positively truncated Gaussians.

\putFrag{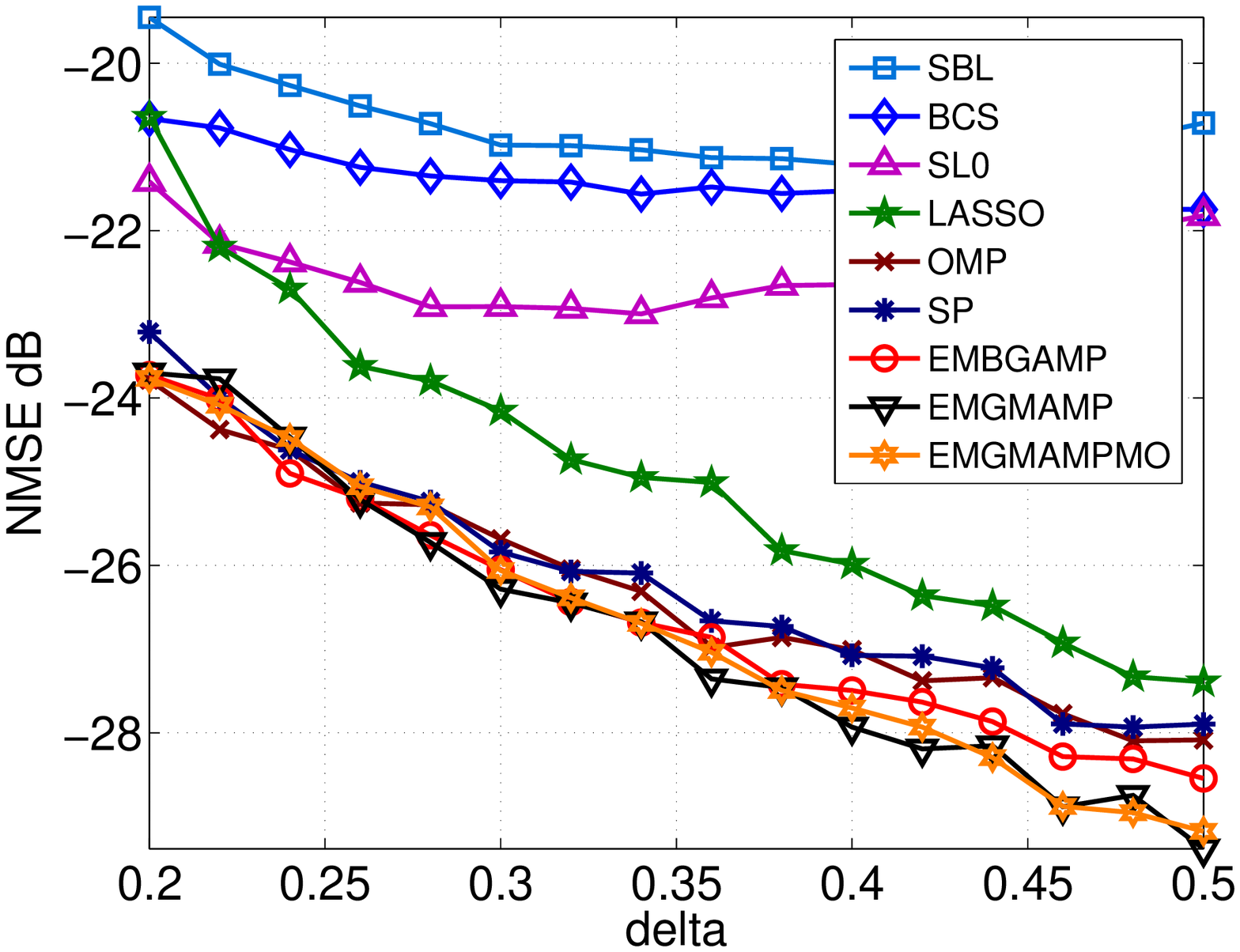}
	{$\NMSE$ versus undersampling ratio $M/N$ for noisy recovery of log-normal signals with location parameter $0$ and scale parameter $1$.}
	{\figsize}
	{\psfrag{delta}[t][t][0.9]{$M/N$} 
	 \psfrag{NMSE dB}[][][0.9]{\sf NMSE [dB]}
         \newcommand{\sz}{0.45}
         \psfrag{EMGMAMPMO}[l][l][\sz]{\sf EM-GM-AMP-MOS}
         \psfrag{EMGMAMP}[l][l][\sz]{\sf EM-GM-AMP}
         \psfrag{EMBGAMP}[l][l][\sz]{\sf EM-BG-AMP}
         \psfrag{BCS}[l][l][\sz]{\sf BCS}
         \psfrag{SBL}[l][l][\sz]{\sf T-MSBL}
         \psfrag{SL0}[l][l][\sz]{\sf SL0}
          \psfrag{SP}[l][l][\sz]{\sf genie SP}
         \psfrag{OMP}[l][l][\sz]{\sf genie OMP}
         \psfrag{LASSO}[l][l][\sz]{\sf genie SPGL1}}

It may be interesting to notice that, with the perfectly sparse signals examined in \figsref{BGNMSE}{BRNMSE}, SL0 and SPGL1 performed relatively poorly, the relevance-vector-machine (RVM)-based approaches (i.e., BCS, T-MSBL) performed relatively well, and the greedy approaches (OMP and SP) performed in-between.
With the heavy-tailed signals in \figsref{STNMSE}{LOGNNMSE}, it is more difficult to see a consistent pattern.
For example, with the Student's-t signal, the greedy approaches performed the worse, the RVM approaches were in the middle, and SL0 and SPGL1 performed very well.
But with the log-normal signal, the situation was very different: the greedy approaches performed very well, SPGL1 performed moderately well, but SL0 and the RVM approaches performed very poorly.

In conclusion, for \emph{all} of the many signal types tested above, the best recovery performance came from EM-GM-AMP and its MOS variant. 
We attribute this behavior to EM-GM-AMP's ability to tune itself to the signal (and in fact the realization) at hand. 

%%%%%%%%%%%%%%%%%%%%%%%%%%%%%%%%%%%%%%%%%%%%%%%%%%%%%%%%%%%%%%%%%%%%%
\subsection{Runtime and Complexity Scaling with $N$}	\label{sec:complexity}

Next we investigated how complexity scales with signal length $N$ by evaluating the runtime of each algorithm on a typical personal computer.
For this, we fixed $K/N = 0.1$, $M/N = 0.5$, $\SNR=25$ dB and varied the signal length $N$.  
\Figref{timelogn} shows the runtimes for noisy recovery of a Bernoulli-Rademacher signal, while \figref{nmselogn} shows the corresponding $\NMSE$s.
In these plots, each datapoint represents an average over $R=50$ realizations.
The algorithms that we tested are the same ones that we described earlier.
However, to fairly evaluate runtime, we configured some a bit differently than before.
In particular, for genie-tuned SPGL1, in order to yield a better runtime-vs-NMSE tradeoff, we reduced the tolerance grid (recall footnote \ref{spl}) to $\sigma^2\in\{0.6,0.8,\dots,1.4\}\times M\psi$ and turned off debiasing.
For OMP and SP, we used the fixed support size $K\lasso = M \rho_\textsf{SE}(\frac{M}{N})$ rather than searching for the size that minimizes $\NMSE$ over a grid of $10$ hypotheses, as before.
Otherwise, all algorithms were run under the suggested defaults, 
with T-MSBL run under \texttt{noise=small}
and EM-GM-AMP run in ``sparse'' mode.

The complexities of the proposed EM-GM-AMP methods are dominated by one matrix multiplication by $\vec{A}$ and $\vec{A}\tran$ per iteration. 
Thus, when these matrix multiplications are explicitly implemented and $\vec{A}$ is dense, the total complexity of EM-GM-AMP should scale as $\mathcal{O}(MN)$.
This scaling is indeed visible in the runtime curves of \figref{timelogn}.
There, $\mathcal{O}(MN)$ becomes $\mathcal{O}(N^2)$ since the ratio $M/N$ was fixed, and the horizontal axis plots $N$ on a logarithmic scale, so that this complexity scaling manifests, at sufficiently large values of $N$, as a line with slope $2$.
\Figref{timelogn} confirms that genie-tuned SPGL1 also has the same complexity scaling, albeit with longer overall runtimes.  
Meanwhile, \figref{timelogn} shows T-MSBL, BCS, SL0, OMP, and SP exhibiting a complexity scaling of $\mc{O}(N^3)$ (under fixed $K/N$ and $M/N$), which results in orders-of-magnitude larger runtimes for long signals (e.g., $N\geq 10^4$).
With short signals (e.g., $N < 1300$), though, 
OMP, SP, SL0, and SPGL1 are faster than EM-GM-AMP.
Finally, \figref{nmselogn} verifies that, for most of the algorithms, the $\NMSE$s are relatively insensitive to signal length $N$ when the undersampling ratio $M/N$ and sparsity ratio $K/M$ are both fixed, although 
the performance of EM-GM-AMP improves with $N$ (which is not surprising in light of AMP's large-system-limit optimality properties \cite{Bayati:TIT:11})
and the performance of BCS degrades with $N$. 

Both the proposed EM-GM-AMP methods and SPGL1 can exploit the case where multiplication by $\vec{A}$ and $\vec{A}\tran$ is implemented using a fast algorithm like the fast Fourier transform (FFT)\footnote{%
  For our FFT-based experiments, we used the complex-valued versions of EM-BG-AMP, EM-GM-AMP, and SPGL1.},
which reduces the complexity to $\mc{O}(N\log N)$, and avoids the need to store $\vec{A}$ in memory---a potentially serious problem when $MN$ is large.
The dashed lines in \figsref{timelogn}{nmselogn} (labeled ``\textsf{fft}'') show the average runtime and $\NMSE$ of the proposed algorithms and SPGL1 in case that $\vec{A}$ was a randomly row-sampled FFT.
As expected, the runtimes are dramatically reduced.
While EM-BG-AMP retains its place as the fastest algorithm, SPGL1 now runs $1.5\times$ faster than EM-GM-AMP (at the cost of $14$ dB higher $\NMSE$).
The MOS version of EM-GM-AMP yields slightly better $\NMSE$, but takes $\approx 2.5$ times as long to run as the fixed-$L$ version.

\putFrag{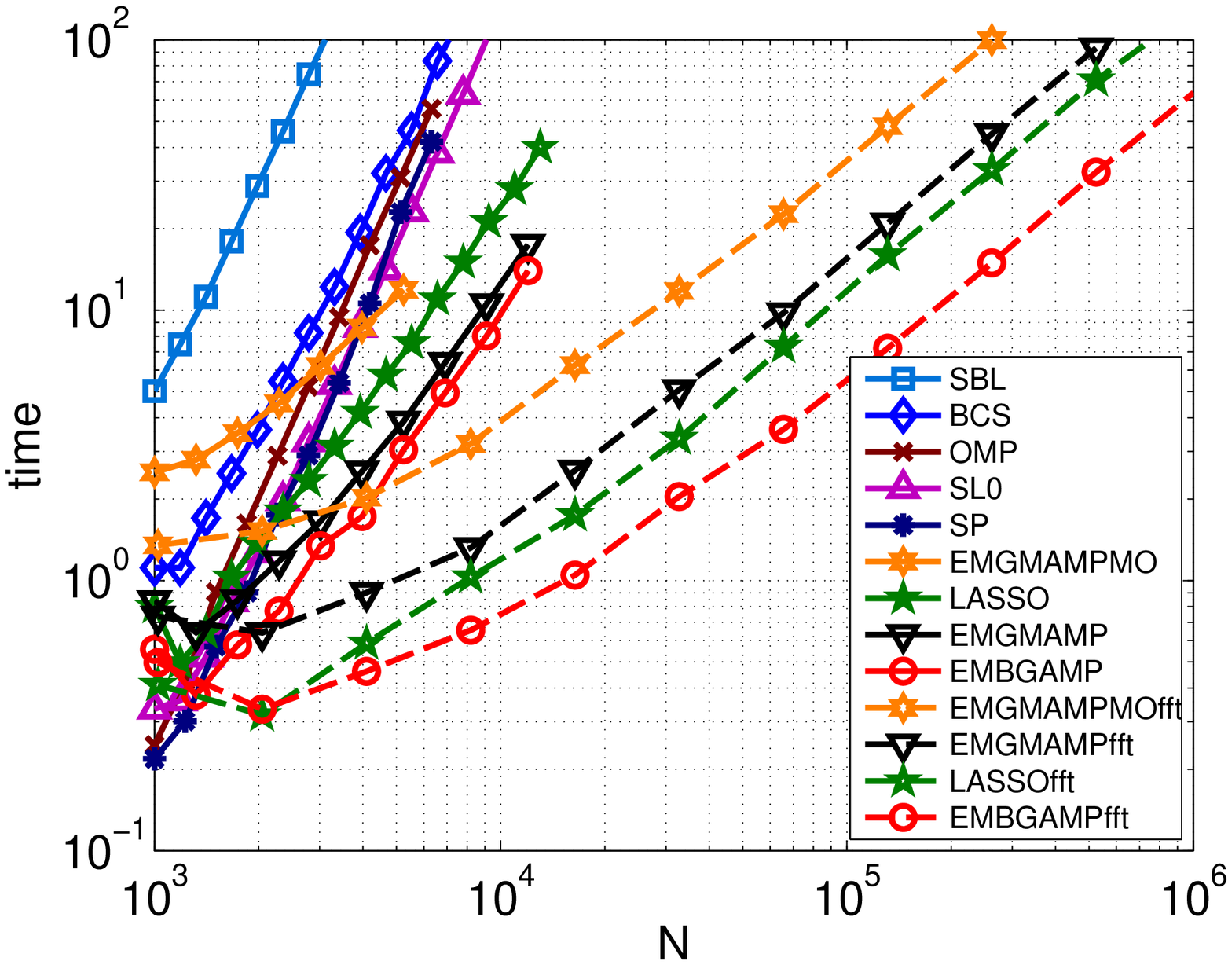}
	{Runtime versus signal length $N$ for noisy recovery of Bernoulli-Rademacher signals.}
	{\figsize}
	{\psfrag{time}[][][0.9]{\sf Runtime [sec]} 
         \newcommand{\sz}{0.40}
         \psfrag{EMGMAMPMOfft}[l][l][\sz]{\sf EM-GM-AMP-MOS fft}
         \psfrag{EMGMAMPMO}[l][l][\sz]{\sf EM-GM-AMP-MOS}
         \psfrag{EMGMAMP}[l][l][\sz]{\sf EM-GM-AMP}
         \psfrag{EMBGAMP}[l][l][\sz]{\sf EM-BG-AMP}
         \psfrag{BCS}[l][l][\sz]{\sf BCS}
         \psfrag{EMGMAMPfft}[l][l][\sz]{\sf EM-GM-AMP fft}
         \psfrag{EMBGAMPfft}[l][l][\sz]{\sf EM-BG-AMP fft}
         \psfrag{LASSOfft}[l][l][\sz]{\sf SPGL1 fft}
         \psfrag{SBL}[l][l][\sz]{\sf T-MSBL}
         \psfrag{SL0}[l][l][\sz]{\sf SL0}
         \psfrag{SP}[l][l][\sz]{\sf SP}
         \psfrag{OMP}[l][l][\sz]{\sf OMP}
         \psfrag{LASSO}[l][l][\sz]{\sf genie SPGL1}
         \psfrag{N}[l][l][0.9]{\sf $N$}}
         
 \putFrag{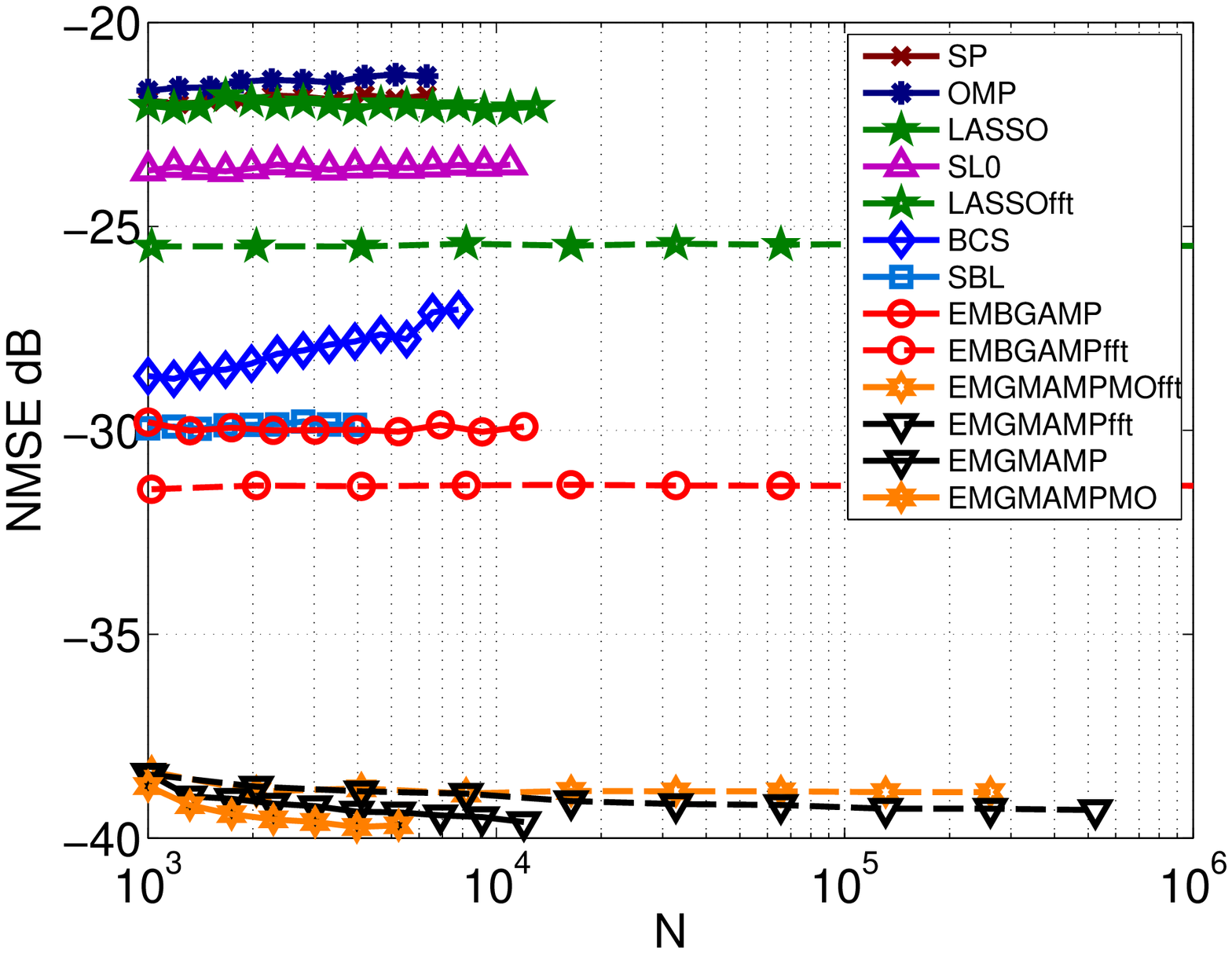}
	{$\NMSE$ versus signal length $N$ for noisy recovery of Bernoulli-Rademacher signals.}
	{\figsize}
	{\psfrag{NMSE dB}[][][0.9]{\sf NMSE [dB]} 
         \newcommand{\sz}{0.40}
         \psfrag{EMGMAMPMOfft}[l][l][\sz]{\sf EM-GM-AMP-MOS fft}
         \psfrag{EMGMAMPMO}[l][l][\sz]{\sf EM-GM-AMP-MOS}
         \psfrag{EMGMAMP}[l][l][\sz]{\sf EM-GM-AMP}
         \psfrag{EMBGAMP}[l][l][\sz]{\sf EM-BG-AMP}
         \psfrag{BCS}[l][l][\sz]{\sf BCS}
         \psfrag{EMGMAMPfft}[l][l][\sz]{\sf EM-GM-AMP fft}
         \psfrag{EMBGAMPfft}[l][l][\sz]{\sf EM-BG-AMP fft}
         \psfrag{LASSOfft}[l][l][\sz]{\sf SPGL1 fft}
         \psfrag{SBL}[l][l][\sz]{\sf T-MSBL}
         \psfrag{SL0}[l][l][\sz]{\sf SL0}
         \psfrag{SP}[l][l][\sz]{\sf SP}
         \psfrag{OMP}[l][l][\sz]{\sf OMP}
         \psfrag{LASSO}[l][l][\sz]{\sf genie SPGL1}
         \psfrag{N}[l][l][0.9]{\sf $N$}}

\subsection{Example: Compressive Recovery of Audio}

As a practical example, we experimented with the recovery of an audio signal from compressed measurements.
The full length-$81920$ audio signal was first partitioned into $T$ blocks $\{\vec{u}_t\}_{t=1}^T$ of length $N$.
Noiseless compressed measurements $\vec{y}_t = \vec{\Phi u}_t \in \Real^M$ were then collected using $M=N/2$ samples per block.
Rather than reconstructing $\vec{u}_t$ directly from $\vec{y}_t$, we first reconstructed\footnote{%
Although one could exploit additional structure among the multiple-timestep coefficients $\{\vec{x}_t\}_{t=1}^T$ for improved recovery (e.g., sparsity clustering in the time and/or frequency dimensions, as well as amplitude correlation in those dimensions) as demonstrated in \cite{Ziniel:SSP:12}, such techniques are outside the scope of this paper.} 
the transform coefficients $\vec{x}_t=\vec{\Psi}\tran \vec{u}_t$, using the (orthogonal) discrete cosine transform (DCT) $\vec{\Psi}\in\Real^{N\times N}$,
 and later reconstructed $\vec{u}_t$ via $\vec{u}_t=\vec{\Psi}\vec{x}_t$.  
Our effective sparse-signal model can thus be written as $\vec{y}_t=\vec{A}\vec{x}_t$ with $\vec{A}=\vec{\Phi\Psi}$.
We experimented with two types of measurement matrix $\vec{\Phi}$: i.i.d zero-mean Gaussian and random selection (i.e., containing rows of the identity matrix selected uniformly at random), noting that the latter allows a fast implementation of $\vec{A}$ and $\vec{A}\tran$.
\tabref{audio} shows the resulting time-averaged $\NMSE$, i.e., $\TNMSE \defn \frac{1}{T} \sum_{t=1}^T ||\vec{u}_t - \vec{\hat{u}}_t||^2/||\vec{u}_t||^2$, and total runtime achieved by the previously described algorithms 
at block lengths $N = 1024, 2048, 4096, 8192$, which correspond to $T = 80, 40, 20,10$ blocks, respectively.  
The numbers reported in the table represent an average over $50$ realizations of $\vec{\Phi}$.
For these experiments, we configured the algorithms as described in \secref{heavy} for the heavy-tailed experiment except that, for genie-SPGL1, rather than using $\psi=0$, we used $\psi=10^{-6}$ for the tolerance grid (recall footnote \ref{spl}) because we found that this value minimized $\TNMSE$
and, for T-MSBL, we used the setting \texttt{prune\_gamma} = $10^{-12}$ as recommended in a personal correspondence with the author.
For certain combinations of algorithm and blocklength, excessive runtimes prevented us from carrying out the experiment, and thus no result appears in the table.

\tabref{audio} shows that, for this audio experiment, the EM-GM-AMP methods and SL0 performed best in terms of $\TNMSE$.
As in the synthetic examples presented earlier, we attribute EM-GM-AMP's excellent $\TNMSE$ to its ability to tune itself to whatever signal is at hand. 
As for SL0's excellent $\TNMSE$, we reason that it had the good fortune of being particularly well-tuned to this audio signal, given that it performed relatively poorly with the signal types used for \figsref{BGNMSE}{SNRNMSE} and \figref{LOGNNMSE}.
From the runtimes reported in \tabref{audio}, we see that, with i.i.d Gaussian $\vec{\Phi}$ and the shortest block length ($N=1024$), genie-OMP is by far the fastest, whereas the EM-GM-AMP methods are the slowest.
But, as the block length grows, the EM-GM-AMP methods achieve better and better runtimes as a consequence of their excellent complexity scaling, and eventually EM-BG-AMP and fixed-$L$ EM-GM-AMP become the two fastest algorithms under test (as shown with i.i.d Gaussian $\vec{\Phi}$ at $N=8192$).
For this audio example, the large-block regime may be the more important, because that is where all algorithms give their smallest $\TNMSE$.
Next, looking at the runtimes under random-selection $\vec{\Phi}$, we see dramatic speed improvements for the EM-GM-AMP methods and SPGL1, which were all able to leverage Matlab's fast DCT.
In fact, the total runtimes of these four algorithms \emph{decrease} as $N$ is increased from $1024$ to $8192$. 
We conclude by noting that EM-BG-AMP (at $N=8192$ with random selection $\vec{\Phi}$) achieves the fastest runtime in the entire table while yielding a $\TNMSE$ that is within $1.3$ dB of the best value in the entire table.
Meanwhile, fixed-$L$ EM-GM-AMP (at $N=8192$ with random selection $\vec{\Phi}$) gives $\TNMSE$ only $0.3$ dB away from the best in the entire table with a runtime of only about twice the best in the entire table.
Finally, the best $\TNMSE$s in the entire table are achieved by EM-GM-AMP-MOS (at $N=8192$), which takes $\approx 2.5$ times as long to run as its fixed-$L$ counterpart.

\putTable{audio}{Average $\TNMSE$ (in dB) and total runtime (in seconds) for compressive audio recovery.}{
\begin{tabular}{|@{\,}c@{\,}|@{\,}c@{\,}|@{}r@{\,}|@{}r@{\,\,}||@{\,}r@{\,}|@{\,}r@{\,\,}||@{\,}r@{\,}|@{\,}r@{\,\,}||@{\,}r@{\,}|@{}r@{\,\,}|}
\cline{3-10}
\multicolumn{2}{c}{} &\multicolumn{2}{|c||}{\tiny $N = 1024$} &\multicolumn{2}{|c||}{\tiny $N=2048$} &\multicolumn{2}{|c||}{\tiny $N=4096$} &\multicolumn{2}{|c|}{\tiny $N=8192$} \\ 
\cline{3-10}
\multicolumn{2}{c|}{} & \,\tiny $\TNMSE$ & \scriptsize time &\tiny $\TNMSE$ &\scriptsize time &\tiny $\TNMSE$ &\scriptsize time &\tiny $\TNMSE$ &\scriptsize time \\ \cline{3-10}
\cline{3-10} \hline
\multirow{7}{*}{\begin{sideways}\parbox{23mm}{i.i.d Gaussian $\vec{\Phi}$}\end{sideways}}
&\scriptsize EM-GM-AMP-MOS &\textbf{-17.3} &\,468.9 &\textbf{-18.3} &487.2 &\textbf{-21.0} & 967.9 &\textbf{-21.8} &2543\\  \cline{2-10}
&EM-GM-AMP &-16.9 &159.2 &-18.0 &213.2 &-20.7 & 434.0 &-21.4 &1129\\  \cline{2-10}
&EM-BG-AMP &-15.9 &115.2 &-17.0 &174.1 &-19.4 &430.2 &-20.0 &\textbf{1116}\\  \cline{2-10} 
&SL0 &-16.8 &41.6 &-17.9 &128.5 &-20.6 & 629.0 &-21.3 &2739\\  \cline{2-10}
&genie SPGL1 &-14.3 &90.9 &-16.2 &200.6 &-18.6 &514.3 &-19.5 &1568\\  \cline{2-10}
&BCS & -15.0 &67.5 &-15.8 &149.1 &-18.4 &\textbf{428.0} &-18.8 &2295\\  \cline{2-10}
&T-MSBL &-16.3 &1.2e4 & -- &-- &-- &--  &-- &-- \\ \cline{2-10}
&genie OMP &-13.9 &\textbf{20.1}&-14.9 &\textbf{109.9}&-17.6 &527.0 &-- &--\\ \cline{2-10} 
&genie SP &-14.5 & 87.7 &-15.5 &305.9 &-18.0 &1331 &-- &--\\ \hline \hline

\multirow{7}{*}{\begin{sideways}\parbox{24mm}{random selection $\vec{\Phi}$}\end{sideways}}
&\scriptsize EM-GM-AMP-MOS &-16.6 &233.0 &-17.5 &136.1 &\textbf{-20.5} & 109.6 &\textbf{-21.6} &93.9\\  \cline{2-10}
&EM-GM-AMP &\textbf{-16.7} &56.1 &\textbf{-17.7} &43.7 &\textbf{-20.5} &38.0 &-21.5 &37.8 \\  \cline{2-10}
&EM-BG-AMP &-16.2 &29.6 &-17.2 &\textbf{22.3} &-19.7 &\textbf{19.4} &-20.5 &\textbf{18.0}\\  \cline{2-10} 
&SL0 &\textbf{-16.7} &35.7 &-17.6 &119.5 &-20.4 &597.8 &-21.2 &2739\\  \cline{2-10}
&genie SPGL1 &-14.0 &34.4 &-15.9 &24.5&-18.4 &21.7 &-19.7 &19.6\\  \cline{2-10}
&BCS & -15.5 &60.5 &-16.1 &126.2 &-19.4 &373.8 &-20.2 &2295\\  \cline{2-10}
&T-MSBL &-15.5 &1.2e4 &-- &-- &-- &-- & -- &-- \\ \cline{2-10}
&genie OMP &-15.1 &\textbf{20.1} &-15.7 &106.8 &-18.9 &506.0 &-- &--\\ \cline{2-10} 
&genie SP &-15.2& 104.5 &-16.1 &395.3&-18.7&1808 &-- &--\\ \hline
\end{tabular}
}

%%%%%%%%%%%%%%%%%%%%%%%%%%%%%%%%%%%%%%%%%%%%%%%%%%%%%%%%%%%%%%%%%%%%%%%%%%%%%
\section{Conclusions}				\label{sec:conc}

Those interested in practical compressive sensing face the daunting task of choosing among literally hundreds of signal reconstruction algorithms (see, e.g., \cite{riceweb}).
In testing these algorithms, they are likely to find that some work very well with particular signal classes, but not with others.
They are also likely to get frustrated by those algorithms that require the tuning of many parameters.
Finally, they are likely to find that some of the algorithms that are commonly regarded as ``very fast'' are actually very slow in high-dimensional problems. 
Meanwhile, those familiar with the theory of compressive sensing know that the workhorse LASSO is nearly minimax optimal, and that its phase transition curve is robust to the nonzero-coefficient distribution of sparse signals.
However, they also know that, for most signal classes, there is a large gap between the MSE performance of LASSO and that of the MMSE estimator derived under full knowledge of the signal and noise statistics \cite{Wu:TIT:12}.
Thus, they may wonder whether there is a way to close this gap by designing a signal reconstruction algorithm that \emph{both learns and exploits} the signal and noise statistics. 

With these considerations in mind, we proposed an empirical Bayesian approach to compressive signal recovery that merges two powerful inference frameworks: expectation maximization (EM) and approximate message passing (AMP).
We then demonstrated---through a detailed numerical study---that our approach, when used with a flexible Gaussian-mixture signal prior, achieves a state-of-the-art combination of reconstruction error and runtime on a very wide range of signal and matrix types in the high-dimensional regime.
However, certain non-zero-mean and super-Gaussian sensing matrices give our AMP-based method trouble. Making AMP robust to these matrices remains a topic of importance for future research.

%%%%%%%%%%%%%%%%%%%%%%%%%%%%%%%%%%%%%%%%%%%%%%%%%%%%%%%%%%%%%%%%%%%%%%%%%%%%%
\bibliographystyle{ieeetr}
\bibliography{macros_abbrev,books,misc,sparse,machine}

%%%%%%%%%%%%%%%%%%%%%%%%%%%%%%%%%%%%%%%%%%%%%%%%%%%%%%%%%%%%%%%%%%%%%%%%%%%%%
\begin{IEEEbiography}[{\includegraphics[width=1.0in,height=1.25in,trim=12mm 15mm 15mm 15mm,clip,keepaspectratio]{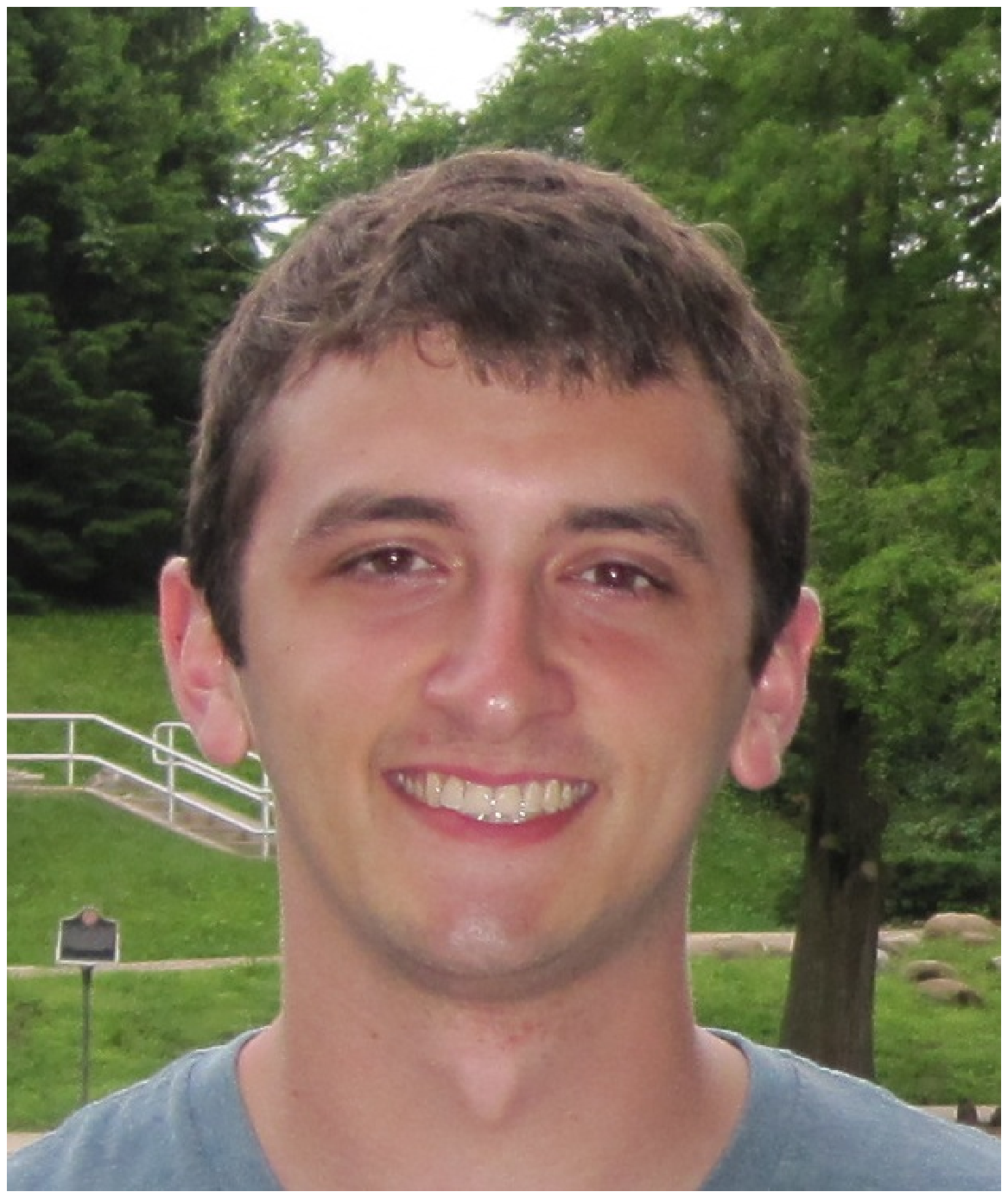}}]{Jeremy Vila} received the B.S.\ degree in Electrical and Computer Engineering from the Ohio State University in 2010.
 He is currently a Ph.D.\ student in the Information Processing Systems Lab in
the Department of Electrical and Computer Engineering at OSU.  His primary
research interests include compressive sensing, statistical signal
processing, and machine learning.
\end{IEEEbiography}

\begin{IEEEbiography}[{\includegraphics[width=1.0in,height=1.25in,trim=10mm 10mm 10mm 10mm,clip,keepaspectratio]{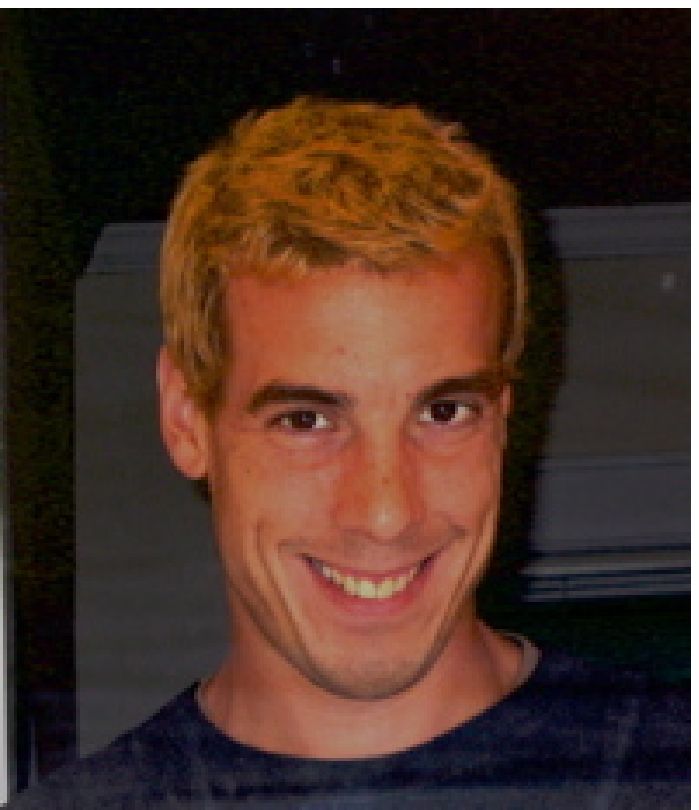}}]{Philip Schniter}
received the B.S.\ and M.S.\ degrees in Electrical Engineering from the University of Illinois at Urbana-Champaign in 1992 and 1993, respectively, and the Ph.D.\ degree in Electrical Engineering from Cornell University in Ithaca, NY, in 2000. 

From 1993 to 1996 he was employed by Tektronix Inc.\ in Beaverton, OR as a
systems engineer. 
After receiving the Ph.D.\ degree, he joined the Department of Electrical and Computer Engineering at The Ohio State University, Columbus, where he is currently 
a Professor and a member of the Information Processing Systems (IPS) Lab.
In 2008-2009 he was a visiting professor at Eurecom, Sophia Antipolis, France,
and Sup{\'e}lec, Gif-sur-Yvette, France.

In 2003, Dr.\ Schniter  received the National Science Foundation CAREER Award.
His areas of interest currently include statistical signal processing, wireless communications and networks, and machine learning.
\end{IEEEbiography}

\end{document}